\documentclass[a4paper,11pt]{article}
\usepackage{jheppub} % for details on the use of the package, please
\usepackage{graphicx} % Required for inserting images
\usepackage{amssymb}
\usepackage{amsfonts}
\usepackage{dsfont}
\usepackage{relsize}
\usepackage{mathtools}
\usepackage{array}
\usepackage{comment}
\usepackage{cases}
\usepackage{empheq}
\usepackage{rotating}
\usepackage{bbold,amsfonts}
\allowdisplaybreaks
\usepackage{scalerel}
\usepackage{amsmath,bm}
\usepackage{amsthm}
\usepackage{physics}
\usepackage[compat=1.1.0]{tikz-feynman}
\usepackage[T1]{fontenc} % if needed
\usepackage{tensor}
\usepackage[export]{adjustbox}
\usepackage{slashed}
\usepackage{stackrel,amssymb}
\usepackage{tikz}
\usepackage{mathtools}
\usetikzlibrary{arrows.meta}
\tikzset{>={Latex[scale=1.1]}}
\usetikzlibrary{arrows.meta,
	bending,
	decorations.markings, decorations.text}
\newcommand{\ee}{\mathrm{e}}
\newcommand{\ii}{\mathrm{i}}

\makeatletter
\newcommand*{\letterdef@}{}
\newcommand*{\letterdef}[3]{%
	\def\letterdef@##1{\expandafter\newcommand\csname #1\endcsname{#2{##1}}}%
	\@tfor\@tempa :=#3\do{\expandafter\letterdef@\expandafter{\@tempa}}}
\makeatother
\letterdef{c#1} {\mathcal}{ABCDEFGHIJKLMNOPQRSTUVWXYZ} % \cX = \mathcal{X}
\letterdef{rm#1}{\mathrm} {dDeimM} % \rmX = \mathrm{X} for X in {dDmM}
\newcommand{\D}{{\scriptscriptstyle{\mathbf{D}}}}
\newcommand{\E}{{\scriptscriptstyle{\mathbf{E}}}}

\newcommand{\Dhat}{{\scriptscriptstyle{\widehat{\mathbf{D}}}}}

\newcommand{\NN}{{\scriptscriptstyle{\,\mathcal{N}=4}}}
\newcommand{\adj}{{\scriptscriptstyle{\mathrm{adj}}}}
\newcommand{\fund}{{\scriptscriptstyle{\mathrm{F}}}}
\newcommand{\asymm}{{\scriptscriptstyle{\mathrm{A}}}}
\newcommand{\symm}{{\scriptscriptstyle{\mathrm{S}}}}
\allowdisplaybreaks
\newdimen\tableauside\tableauside=1.0ex
\newdimen\tableaurule\tableaurule=0.4pt
\newdimen\tableaustep
\def\phantomhrule#1{\hbox{\vbox to0pt{\hrule height\tableaurule
			width#1\vss}}}
\def\phantomvrule#1{\vbox{\hbox to0pt{\vrule width\tableaurule
			height#1\hss}}}
\def\sqr{\vbox{%
		\phantomhrule\tableaustep
		\hbox{\phantomvrule\tableaustep\kern\tableaustep\phantomvrule\tableaustep}%
		\hbox{\vbox{\phantomhrule\tableauside}\kern-\tableaurule}}}
\def\squares#1{\hbox{\count0=#1\noindent\loop\sqr
		\advance\count0 by-1 \ifnum\count0>0\repeat}}
\def\tableau#1{\vcenter{\offinterlineskip
		\tableaustep=\tableauside\advance\tableaustep by-\tableaurule
		\kern\normallineskip\hbox
		{\kern\normallineskip\vbox
			{\gettableau#1 0 }%
			\kern\normallineskip\kern\tableaurule}%
		\kern\normallineskip\kern\tableaurule}}
\def\gettableau#1 {\ifnum#1=0\let\next=\null\else
	\squares{#1}\let\next=\gettableau\fi\next}
\newcommand{\Yfund}{\tableau{1}}
\newcommand{\Ysymm}{\tableau{2}}
\newcommand{\Yasymm}{\tableau{1 1}}

\title{Strong-coupling results in (non-)conformal $\mathcal{N}=2$ theories with fundamental flavors}

\author[a,b]{M. Bill\`o,}
\author[c,b]{A. Lerda,}
\author[d]{and A. Testa\,}

\affiliation[a]{Universit\`a di Torino, Dipartimento di Fisica,\\ Via P. Giuria 1, I-10125 Torino, Italy}
	\affiliation[b]{INFN, Sezione di Torino,	\\Via P. Giuria 1, I-10125 Torino, Italy}
\affiliation[c]{Universit\`a del Piemonte Orientale,
			Dipartimento di Scienze e Innovazione Tecnologica\\
			Viale T. Michel 11, I-15121 Alessandria, Italy}
	\affiliation[d]{Institut de Physique Th\'eorique\footnote{Unit\'e Mixte de Recherche 3681 du CNRS}, Universit\'e Paris Saclay, CNRS,  91191 Gif-sur-Yvette, France}
\emailAdd{marco.billo@unito.it}
\emailAdd{lerda@to.infn.it}
\emailAdd{alessandro.testa@ipht.fr
}

\abstract{
We study a class of four-dimensional $\mathcal{N}=2$ SU($N$) gauge theories with two massless hypermultiplets in the rank-two antisymmetric representation and $0\leq N_\fund\leq 4$ fundamental flavors. These theories are superconformal for $N_\fund=4$ and asymptotically free otherwise. Supersymmetric localization on the four-sphere applies in both cases and leads to matrix-model representations for a broad class of protected observables.
Within this framework, we compute the free energy and the expectation value of the $\frac{1}{2}$-BPS circular Wilson loop in the large-$N$ limit. For any $N_\fund$, we show that the infinite series of non-planar corrections at large 't Hooft coupling $\lambda$ can be resummed in terms of an effective coupling $\widetilde{\lambda}$. In the non-conformal window ($0\leq N_\fund<4$), this provides a first example in which localization leads to explicit results in the planar limit at strong 't Hooft coupling. In the superconformal case ($N_\fund=4$), the large-$N$ expansion in $\widetilde{\lambda}$ suggests a refined AdS/CFT dictionary in which $\widetilde{\lambda}$, rather than $\lambda$, plays the role of the bulk string coupling.
}

\begin{document}
\maketitle
\flushbottom
\section{Introduction and summary of results}
\label{intro}
One of the central challenges in current theoretical physics is to develop analytic control over strongly coupled quantum field theories. A particularly fertile arena to address this problem is provided by four-dimensional gauge theories with extended supersymmetry, where powerful non-perturbative tools are available.

Among these, holographic AdS/CFT duality \cite{Aharony:1999ti} and supersymmetric localization \cite{Pestun:2016zxk} play a pivotal role and often complement each other. On the one hand, localization can extrapolate  strong-coupling predictions which can be compared directly with holographic results. This approach is especially valuable when the holographic answer is not completely fixed, as is the case, for example, of integrated correlators \cite{Binder:2019jwn} whose exact expression obtained via localization can be used to impose non-trivial constraints on holographic amplitudes. On the other hand, holography can shed new light on the reorganization of the gauge-theory degrees of freedom at strong coupling. 

It is therefore of clear interest to enlarge the class of theories where both tools can be simultaneously exploited.
On the holographic side, the paradigmatic example of gauge/gravity duality relates $\mathcal{N}=4$ super Yang--Mills (SYM) theory to Type IIB string theory on $\mathrm{AdS}_5 \times S^5$ \cite{Maldacena:1997re}. However, dual descriptions are  also available  for theories with reduced supersymmetry and even for non-conformal models. On the other hand, localization requires at least $\mathcal{N}=2$ supersymmetry, but not conformal invariance \cite{Pestun:2007rz}, since it is naturally formulated on compact manifolds such as (possibly squashed) four-spheres $S^4$. For superconformal setups, the conformal equivalence between $S^4$ and $\mathbb{R}^4$ naturally relates localization results to flat space, thereby justifying why this technique has so far been applied mostly to $\mathcal{N}=4$ SYM and to $\mathcal{N}=2$ superconformal models.

When conformal symmetry is explicitly broken by mass deformations, localization still reproduces correctly perturbative field-theory computations on $S^4$, but no longer agrees with flat-space results \cite{Belitsky:2020hzs}. A more subtle and interesting situation arises when conformal symmetry is broken at the quantum level by a non-vanishing $\beta$-function, while the matter content remains massless. In this case, the matrix model obtained by localization on a sphere of radius $R$ still captures \emph{part} of the physical information for some flat-space observables within a specific regime of validity, provided that $R$ is identified with their characteristic physical scale and the matrix-model coupling is interpreted as the running coupling $g$ evaluated at that scale. This correspondence has been explicitly verified up to order $g^6$ for the vacuum expectation value of the $\frac{1}{2}$-BPS Wilson loop \cite{Billo:2023igr,Billo:2024hvf,Billo:2024fst} and for chiral/anti-chiral correlators \cite{Billo:2019job,Billo:2025sxr}.

Matrix-model computations are typically much simpler than their quantum field theory counterparts. Moreover, in several superconformal cases, reorganizing these matrix models in a large-$N$ 't Hooft expansion makes it possible to probe the strong-coupling regime \cite{Russo:2012ay,Beccaria:2020hgy}. This is precisely where comparisons with holography are most natural, since the large-$N$ expansion corresponds to the topological world-sheet expansion on the dual side. It is therefore important to identify non-conformal theories for which localization still provides access to strong-coupling regimes and a holographic description may plausibly exist.

In this paper we focus on a specific class of $\mathcal{N}=2$ SYM theories with gauge group $\mathrm{SU}(N)$ and matter consisting of two hypermultiplets in the antisymmetric representation and $N_F$ hypermultiplets in the fundamental. The one-loop $\beta$-function is proportional to $b_0=4-N_F$, so the theory is superconformal for $N_F = 4$ and asymptotically free for $N_F < 4$. These theories admit a Type IIB string realization as a system of $N$ D3-branes in the presence of an O7-plane and four D7-branes \cite{Fayyazuddin:1998fb,Aharony:1998xz,Park:1998zh,Billo:2010uaa}.

In the conformal case, often referred to as the \textbf{D}-theory \cite{Billo:2019fbi}, all four D7-branes sit on top of the O7-plane and the D3-branes. The antisymmetric matter hypermultiplets arise from massless D3/D3 open strings affected by the orientifold projection, while the fundamental hypermultiplets come from massless D3/D7 strings, yielding $N_F = 4$ and hence $b_0 = 0$. The resulting U(4) flavor symmetry originates from the Chan-Paton factors of the D7-branes. Since the dilaton sources of the O7-plane and the D7-branes cancel exactly, the dilaton remains constant. In the near-horizon limit of the D3-branes, the holographic dual geometry is given by Type IIB string theory on $\mathrm{AdS}_5 \times S^5$, modded out by a $\mathbb{Z}_2 \times \mathbb{Z}_2$ orbifold/orientifold action \cite{Ennes:2000fu}. The fixed locus of the orbifold action is $\mathrm{AdS}_5\times {S^1}$ and supports the twisted sector; the fixed point of the orientifold action is instead $\mathrm{AdS}_5\times {S^3}$ and supports the O7 plane and the four D7 branes.

The asymptotically free case with $N_F < 4$, which we refer to as $\widehat{\mathbf{D}}$-theory, may be engineered by displacing $b_0 = 4 - N_F$ of the D7-branes away from the O7-plane and the D3-branes along their common transverse directions. As a result, $b_0$ of the fundamental hypermultiplets acquire a mass $M$ inversely proportional to the brane separation. At energy scales much lower than $M$, these massive fields decouple, leaving an effective theory with $N_F$ massless fundamentals. While the total dilaton source still vanishes globally, it no longer cancels locally: within a region of size $1/M$, the axio-dilaton develops a logarithmic profile proportional to $b_0$ in the transverse directions, reflecting the running of the gauge coupling. Since $b_0$ does not scale with $N$, it is tempting to expect that in the large-$N$ limit these theories admit a holographic description corresponding to a controlled deformation of the conformal background.

In this paper, by significantly extending the analysis of \cite{Beccaria:2021ism}, we show that both the $\mathbf{D}$- and $\widehat{\mathbf{D}}$-theories can be studied efficiently using supersymmetric localization. In both cases the resulting matrix model contains  double-trace and single-trace interactions. Remarkably, the double-trace sector coincides exactly with that of the matrix model associated with the superconformal $\mathcal{N}=2$ theory known as  $\mathbf{E}$-theory \cite{Billo:2019fbi}, whose matter transforms in the direct sum of a symmetric and an antisymmetric representation of the gauge group.

The $\mathbf{E}$-theory matrix model has been extensively studied \cite{Beccaria:2020hgy,Beccaria:2021vuc,Beccaria:2021hvt,Billo:2022xas,Billo:2022gmq,Billo:2022fnb,Beccaria:2022ypy,Billo:2023kak,Beccaria:2023kbl}; in particular, its free energy and the vacuum expectation value of the $\frac{1}{2}$-BPS Wilson loop have been recently analyzed in the large-$N$ limit and at strong 't~Hooft coupling $\lambda$ in \cite{Beccaria:2023kbl}. A key ingredient of that analysis, also central  to the present work, is a Hubbard--Stratonovich transformation that introduces gauge-invariant dynamical variables $K_j$ dual to single-trace operators of odd dimension. The resulting effective description for the $K_j$'s admits a cumulant expansion whose vertices are connected multi-trace correlators in the Gaussian matrix model.

Such connected Gaussian correlators have been considered in both the physics and mathematics literature (see \cite{Beccaria:2023kbl} and references therein). They depend on the parity of the trace powers and admit a systematic topological expansion in powers of $1/N$, with coefficients given by quasi-polynomials in the trace orders. Determining these quasi-polynomials explicitly is a highly non-trivial combinatorial problem, and only partial results are currently available\,\footnote{In Appendix~\ref{sec:GC} we extend the existing results by deriving new cases relevant for the present analysis.}. In \cite{Beccaria:2023kbl} this approach was used to study the large-$N$ strong-coupling regime of the $\mathbf{E}$-theory; here we show that the same framework can be adapted successfully to the $\widehat{\mathbf{D}}$-theory, which includes the conformal $\mathbf{D}$-theory as a particular case.

\subsection{Outline and summary of results}
\label{subsec:summ-res}
In Section~\ref{sec:matrix-model} we present the matrix-model description of the $\widehat{\mathbf{D}}$-theory and show that it can be reformulated as an effective model for a set of gauge-invariant variables $K_j$. The interaction vertices of this effective theory are determined by multi-trace correlators in a free Gaussian matrix model. The quasi-polynomial structure of these correlators makes it possible to efficiently analyze their dependence on $N$ and on the 't Hooft coupling $\lambda$, and applies equally well to conformal and non-conformal cases.

In Section~\ref{secn:freeenergy+WL} we focus on two observables of the $\widehat{\mathbf{D}}$-theory: the free energy $F_\Dhat$ and
the vacuum expectation value of a $\frac{1}{2}$-BPS Wilson loop $\langle\mathcal{W}\rangle_\Dhat$. Exploiting the properties of the Gaussian correlators, we derive their large-$N$ expansion. The strong-coupling behavior of these observables at each order in $1/N$ is then analyzed in Sections~\ref{secn:free energy} and \ref{secn:Wilson loop} using Mellin-Barnes techniques. In particular, we show that in the regime
\begin{align}
\begin{cases}
    \!\!&1\ll\lambda\ll N \quad\mbox{if}~~b_0=0~,\\[2mm]
    \!\!&1\ll\lambda\ll\lambda\log\lambda\ll N\quad\mbox{if}~~b_0\not=0~,
\end{cases}
\label{regimes_intro}
\end{align}
the leading large-$\lambda$ contributions, which originate from the single-trace interactions of the matrix model, can be resummed to all orders in $1/N$. As a result, both observables can be expressed in terms of an effective rescaled coupling $\widetilde{\lambda}$ defined by
\begin{subequations}
\begin{numcases}
{\widetilde{\lambda}=}
~\frac{\lambda}{1+\frac{\log 2}{2\pi^2 N}\lambda}
& \text{if } $b_0 = 0$~,\label{tildelambdaconf_intro}
\\[3mm]
~\frac{\lambda}{1+\frac{b_0}{16\pi^2 N}\,\lambda\log\lambda}
& \text{if } $b_0 \neq 0$~. \label{tildelambdanonconf_intro}
\end{numcases}\label{lambdatilde_intro}%
\end{subequations}
The condition $g^2=\lambda/N\ll 1$ implied by (\ref{regimes_intro}) is precisely the regime in which, based on the results of \cite{Billo:2023igr,Billo:2024hvf,Billo:2024fst,Billo:2019job,Billo:2025sxr}, the matrix-model description on $S^4$ remains reliable even when the $\beta$-function is non-vanishing.

\paragraph{The non-conformal case:}
When $b_0\not=0$, we show that at leading order in the strong-coupling expansion the free energy and the  Wilson loop expectation value behave as
\begin{align}
    \label{Fnc0}
        F_\Dhat & \,\underset{\lambda \rightarrow \infty}{\sim}~-N^2\,\frac{\log\widetilde{\lambda}}{2} + \ldots\\[2mm]
         \label{WdisWh00}
        \big\langle \cW \big\rangle_\Dhat & ~\underset{\lambda \rightarrow \infty}{\sim}~
        \frac{2\,I_1\big(\!\scaleobj{0.9}{\sqrt{\widetilde\lambda}}\,\big)}{\scaleobj{0.9}{\sqrt{\widetilde\lambda}}}+ \ldots
        = \sqrt{\frac{2}{\pi}}\, \frac{\ee^{\sqrt{\widetilde{\lambda}}}}{\widetilde{\lambda}^{3/4}}\bigg(1 - \frac 38 
        \frac{1}{\scaleobj{0.9}{\sqrt{\widetilde\lambda}}} + \ldots\bigg) +\ldots
\end{align}
where $I_\nu$ denotes the modified Bessel function of the first kind and $\widetilde{\lambda}$ is given in (\ref{tildelambdanonconf_intro}). These expressions are valid up to exponentially suppressed terms $O\big(\rme^{-\scaleobj{0.9}{\sqrt{\widetilde\lambda}}}\big)$, and coincide with the planar limit of the free energy and the Wilson loop expectation value in $\mathcal{N}=4$ SYM once written in terms of the rescaled coupling $\widetilde{\lambda}$. Despite their compact form, these results are new and highly non-trivial, since they arise from an all-order resummation of the single-trace contributions in the matrix model of the $\widehat{\mathbf{D}}$-theory, as shown explicitly in Sections~\ref{subscn:freelead} and~\ref{secn:Wilson loop}.

\paragraph{The conformal case:} For $b_0=0$, the rescaling (\ref{tildelambdaconf_intro}), which can also be viewed as a finite renormalization of the Yang-Mills coupling $g^2$, coincides with the one proposed in \cite{Beccaria:2022kxy,Chester:2025ssu} for the $\mathcal{N}=2$ Sp($2N$) theory with one antisymmetric and 4 fundamental hypermultiplets. The same rescaling (\ref{tildelambdaconf_intro}) has also appeared recently in studies of the modular properties of integrated correlators \cite{DeLillo:2025stg} and of the Bremsstrahlung function in the \textbf{D}-theory \cite{DeLillo:2026scl}.
In this case, the strong-coupling analysis can be extended in a systematic way to several subleading orders and all contributions arising from single-trace interactions can be resummed in terms of the rescaled coupling (\ref{tildelambdaconf_intro}).
In Section~\ref{subsecn:subleading} we explicitly carry out this analysis for the free energy of the \textbf{D}-theory and obtain
\begin{align}
    F_\D&\underset{\lambda \rightarrow \infty}{\sim}-N^2\,\frac{\log \widetilde{\lambda}}{2}-N\,\frac{\log \widetilde{\lambda}}{4}+\bigg(\frac{\sqrt{\widetilde{\lambda}}}{8}+\frac{3\log \widetilde{\lambda}}{32}\bigg)+
    \frac{1}{N}\frac{\sqrt{\widetilde{\lambda}}}{64}\notag\\[2mm]
&\qquad~-\frac{1}{N^2}\bigg(\frac{\widetilde{\lambda}^{3/2}}{2048}+\frac{3\widetilde{\lambda}}{2048}-\frac{(9-16\log 2)\sqrt{\widetilde{\lambda}}}{2048}\bigg)+\ldots~
\label{FDstrong_intro}
\end{align}
up to exponentially suppressed contributions $O\big(\rme^{-\scaleobj{0.9}{\sqrt{\widetilde\lambda}}}\big)$. It is interesting to observe that the first two terms can be written as
\begin{align}
    -2\Big(\frac{N^2}{4}+\frac{N}{8}\Big)\log\widetilde{\lambda}
    \label{a-anomaly}
\end{align}
where the expression in parentheses precisely reproduces the $N$-dependent part of the $a$-anomaly coefficient\,\footnote{As explained in \cite{Beccaria:2021ism}, for $\mathcal{N}=2$ theories there is no \emph{a priori} reason to expect that the coefficient of the logarithmic term in the strong-coupling expansion of the free energy be proportional to the $a$-anomaly. However, the fact that the $N$-dependent part of $a$ is correctly obtained is a remarkable fact, see also \cite{Blau:1999vz}.\label{footnote_intro}} of the \textbf{D}-theory, given by $a=\frac{N^2}{4}+\frac{N}{8}-\frac{5}{24}$.
This behavior closely parallels that of the $\mathcal{N}=2$ Sp$(2N)$ model studied in \cite{Beccaria:2022kxy}, where, however, the large-$N$ strong-coupling expansion of the free-energy written in terms of the rescaled coupling $\widetilde{\lambda}$ terminates after finitely many terms due to the absence of double-trace interactions in the matrix model.

\paragraph{Comments on the holographic dual:}
The fact that the large-$N$ strong-coupling expansion of the free energy in the \textbf{D}-theory can be written entirely in terms of the rescaled coupling $\widetilde{\lambda}$ naturally suggests adopting it as the appropriate 't Hooft coupling in the holographic dictionary \cite{Aharony:1999ti,Maldacena:1997re}\,\footnote{A similar observation was done in \cite{Beccaria:2022kxy} for the Sp($2N$) theory.}. This leads us to define the string coupling $g_s$ and the string tension $T$, measured in units of the AdS radius $L$, as
\begin{align}
8\pi g_s=\frac{\widetilde{\lambda}}{N}~,\qquad
2\pi T=\frac{L^2}{\alpha^\prime}=\sqrt{\widetilde{\lambda}}~,
\label{dictionary}
\end{align}
which imply\,\footnote{Notice that these relations differ by some factors of two with respect to the familiar ones of $\mathcal{N}=4$ SYM. These factors are due to the fact that the dual background geometry associated to the \textbf{D}-theory is not AdS$_5\times S^5$ but a projection of it in which $S^5$ is replaced by $S^{\prime5}$ whose volume is $\frac{1}{2}$ of that of $S^5$. Therefore, using the general formula $L^4=4\pi g_s N\frac{\mathrm{vol(S^5)}}{\mathrm{vol}(M_5)}$, valid for a geometry of the type AdS$_5\times M_5$, we find (\ref{L4}) for $M_5=S^{\prime5}$.}
\begin{align}
    L^4=8\pi g_sN\alpha^{\prime\,2}~.
    \label{L4}
\end{align}
The quantity $\tau_2\equiv\frac{8\pi N}{\widetilde{\lambda}}=\frac{1}{g_s}$ can be identified with the imaginary part of the complexified \emph{infrared} coupling $\tau$ which naturally transforms under the SL$(2,\mathbb{Z})$ duality group and differs from its ultraviolet counterpart $\tau_{\scriptscriptstyle{\mathrm{UV}}}$, as discussed in \cite{Billo:2024ftq} for the \textbf{D}-theory and in \cite{Behan:2023fqq} for the closely related Sp($2N$) theory.

In terms of the variables defined in (\ref{dictionary}), the leading terms of (\ref{FDstrong_intro}) take the form
\begin{align}
    F_\D&\sim-\frac{\pi^2 \,T^4\,\log (2\pi T)}{4g_s^2}-\frac{\pi\, T^2\,\log(2\pi T)}{4g_s}+\frac{\pi\,T}{4}+\frac{3\log(2\pi T)}{16}+O(g_s)~.
\label{FDstronghol}
\end{align}
This expansion exhibits the structure expected from the holographic correspondence. The leading $1/g_s^2$-term can be understood in close analogy with the $\mathcal{N}=4$ case as arising from the tree-level (sphere) string effective action $\frac{T^4}{16\pi^3g_s^2}\frac{1}{L^8}\int\! d^{10}x \,\sqrt{g}\,(R+\ldots)$, supplemented by a suitable topological boundary term \cite{Kurlyand:2022vzv}. Evaluating this action on the AdS$_5\times S^{\prime5}$ background and regulating the AdS volume as in \cite{Russo:2012ay}, reproduces the leading term in (\ref{FDstronghol}).

The subleading $1/g_s$-term has a structure consistent with the tree-level (disk) effective action of the D7-branes and orientifold O7-planes. In fact, this action contains curvature-squared terms of the schematic form $\frac{T^2}{16\pi^3g_s}\frac{1}{L^4}\int\! d^8x\, \sqrt{g}\,(R\cdot R+\ldots)$, see for example \cite{Bachas:1999um,Guralnik:2004ve}. Upon integration over AdS$_5\times S^3$ and after regulating the AdS volume as above, these terms yield contributions proportional to $\frac{\pi\, T^2\,\log(2\pi T)}{g_s}$, in agreement with (\ref{FDstronghol}). The precise numerical coefficient, however, cannot be fixed unambiguously at present, since additional couplings, such as those involving the self-dual Ramond-Ramond five-form, may also contribute to the same structure. Owing to the current lack of a complete characterization of these terms in the world-volume action, a holographic determination of the coefficient remains an open problem\,\footnote{See also \cite{Beccaria:2022kxy}.}.

The $g_s^0$-terms in (\ref{FDstronghol}) represent genuine new predictions of the localization computation. In the holographic dual description, they should arise from quantum corrections associated to closed-string loops (torus diagrams) and open-string loops (annulus diagrams). It would be very interesting to verify these contributions directly through a dual holographic calculation. Even more intriguing questions are whether or not the holographic dictionary (\ref{dictionary}) continues to apply in the non-conformal $\widehat{\mathbf{D}}$-theory and how the rescaled coupling $\widetilde{\lambda}$ defined in (\ref{tildelambdanonconf_intro}) should be interpreted from the holographic dual perspective in this case. Since localization provides a firm prediction for the leading strong-coupling behavior of the free energy in the non-conformal case, given in (\ref{Fnc0}), these questions appear to be accessible. However, we leave their investigation to future work.

In Section~\ref{secn:concl} we present our conclusions. Finally, a number of technical details and auxiliary computations, which are useful to reproduce the results presented in the main text, are collected in the appendices.

\section{Matrix-model techniques for \texorpdfstring{$\mathcal{N}=2$}{} SYM theories}
\label{sec:matrix-model}
Exploiting supersymmetric localization \cite{Pestun:2007rz}, the partition function $\mathcal{Z}$ of an $\mathcal{N}=2$ SYM theory with gauge group SU($N$) and coupling $g$, formulated on a four-sphere $S^4$ of radius $R$, can be reduced to a finite-dimensional integral over a matrix $a\in \mathfrak{su}(N)$:
\begin{align}
    \mathcal{Z}= \int \mathcal{D}a~\rme^{-\frac{8\pi^2}{g^2}\,\tr (Ra)^2}\,\big|Z_{\mathrm{1-loop}}\,Z_{\mathrm{inst}}\big|^2~.
    \label{Z}
\end{align}
Here $Z_{\mathrm{1-loop}}$ encodes the Gaussian fluctuations around the localization locus, while $Z_{\mathrm{inst}}$ captures the instanton contributions\,\footnote{As explained in Appendix A of \cite{Beccaria:2021ism}, the partition function $\mathcal{Z}$ contains also an overall coupling-independent normalization factor that accounts for the $a$-anomaly coefficient of the theory. Being coupling-independent, this factor is irrelevant in our discussion, and thus we have omitted it. Our conventions for the trace normalization and the integration measure appearing in (\ref{Z}) are specified below, see (\ref{tracenorm}) and (\ref{measurenorm}).}.
In the large-$N$ limit with fixed 't~Hooft coupling $\lambda=Ng^2$, instanton effects are exponentially suppressed and can be consistently neglected, allowing us to set $Z_{\mathrm{inst}}=1$. Note that these effects remain negligible even when we take the strong-coupling limit in $\lambda$ with $g^2=\lambda/{N}\ll1$. This is the regime considered in this paper.

The one-loop contribution can be expressed as
\begin{align}
    \big|Z_{\mathrm{1-loop}}\big|^2 = \rme^{-S_{\mathrm{int}}(Ra)}~,
\end{align}
thereby defining an interaction action whose explicit form depends on the matter content of the theory.
For massless hypermultiplets in a representation $\mathcal{R}$ of SU($N$), one finds \cite{Billo:2019fbi,Beccaria:2020hgy}
\begin{align}
	\label{eq:interaction action}
	S_{\mathrm{int}}(Ra) = \Tr_{\scriptscriptstyle{\mathcal{R}}}\log H(Ra)-\Tr_\adj \log H(Ra)~\equiv~\Tr_{\scriptscriptstyle{\mathcal{R}}}^\prime\log  H(Ra)~,
\end{align}
where $H(x)$ is defined in terms of the Barnes-$G$ function as
\begin{align}
    H(x)= G(1+\ii x) \,G(1-\ii x)\, \ee^{-(1+\gamma)x^2}
	=  \prod_{n=1}^{\infty}\left(1+\frac{x^2}{n^2}\right)^n \ee^{-\frac{x^2}{n}}~,
\end{align}
with $\gamma$ denoting the Euler-Mascheroni constant.

Using the small-$x$ expansion of $H(x)$, the interaction action (\ref{eq:interaction action}) admits the series representation
\begin{align}
 	\label{eq:Sint_exp}
S_{\mathrm{int}}(Ra) = \sum_{p=1}^{\infty}(-1)^{p}\,\frac{\zeta(2p+1)}{p+1}\Tr_{\scriptscriptstyle{\mathcal{R}}}^\prime (Ra)^{2p+2}~.
\end{align} 
Importantly, these results hold for $\mathcal{N}=2$ SYM theories with arbitrary massless matter, including models with a non-vanishing $\beta$-function. In this case,  the coupling $g$ in (\ref{Z}) must be interpreted as the running coupling evaluated at the scale $1/R$, namely \cite{Pestun:2007rz,Billo:2023igr,Billo:2024hvf,Billo:2024fst,Billo:2025sxr}
\begin{align}
    \label{runningdef}
        \frac{1}{g^2}=\frac{1}{g_\star^2}-\frac{b_0}{8\pi^2}\log(\mu R)~.
\end{align}
In this expression $g_\star$ is the coupling evaluated at the renormalization scale $\mu$ and $b_0$ is the one-loop coefficient of the $\beta$-function. With our conventions, $b_0>0$ corresponds to asymptotically free theories and is given by
\begin{align}
    b_0=2\big(N-i_\mathcal{R}\big) \ ,
    \label{b0def}
\end{align}
where $i_\mathcal{R}$ is the Dynkin index of the representation $\mathcal{R}$.
The presence of a running coupling implies the existence of a strong-coupling scale $\Lambda$ defined by
\begin{equation}
    R \Lambda =  \ee^{-\frac{8\pi^2 }{b_0 g^2}}
\end{equation} 
at which $g$ diverges. Since in this work we restrict to the small-$g$ regime where $R\Lambda$ is exponentially suppressed, all contributions of the form $(R\Lambda)^n$ can be consistently neglected, just like instantons.

Upon rescaling the matrix $a$ as 
\begin{align}
    a\to\sqrt{\frac{\lambda}{8\pi^2N}}\,\frac{{a}}{R}~,
    \label{rescaling}
\end{align}
the partition function (\ref{Z}) becomes
\begin{align}
    \mathcal{Z}= \mathcal{Z}_\NN\,\int\!\mathcal{D}{a}~\rme^{-\tr a^2\,-S_{\mathrm{int}}\big(\sqrt{\frac{\lambda}{8\pi^2 N}}a\big)}\ ,
    \label{Zresc}
\end{align}
where $\mathcal{Z}_\NN$ is the partition function of $\mathcal{N}=4$ SYM with coupling $\lambda$, arising from the Jacobian of the rescaling, while the interaction action takes the form
\begin{align}
S_{\mathrm{int}}\Big(\sqrt{\frac{\lambda}{8\pi^2 N}}a\Big)=\sum_{p=1}^{\infty}(-1)^{p}\,\Big(\frac{\lambda}{8\pi^2N}\Big)^{p+1}\,\frac{\zeta(2p+1)}{p+1}\Tr_{\scriptscriptstyle{\mathcal{R}}}^\prime a^{2p+2}~.
    \label{SintR}
\end{align}
In the following, we adopt the full Lie-algebra formulation introduced in \cite{Billo:2017glv} and write $a=a^b\,T^b$ ($b=1,\ldots,N^2-1)$, where 
the generators $T^b$ of $\mathfrak{su}(N)$ in the fundamental representation are normalized as
\begin{align}
    \tr T^b\,T^c=\frac{1}{2}\,\delta^{bc}~,
    \label{tracenorm}
\end{align}
and the integration measure is chosen as
\begin{align}
    \mathcal{D}a=\prod_{b=1}^{N^2-1}\frac{da^b}{\sqrt{2\pi}}~,
    \label{measurenorm}
\end{align}
so that $\int \mathcal{D}a~\rme^{-\tr a^2}=1$.

\subsection{Special classes of \texorpdfstring{$\mathcal{N}=2$}{} theories}
For a generic matter representation $\mathcal{R}$, the explicit evaluation of the matrix model (\ref{Zresc}) is highly nontrivial due to the presence of the primed trace in the interaction potential (\ref{SintR}). However, substantial simplifications occur when the hypermultiplet representation takes the form
\begin{align}
\mathcal{R}=N_\adj\, \big(\mathrm{adjoint}\big)\, \oplus\, N_\fund ~\big(\Yfund\big)\, \oplus\,N_\asymm~\Big(\Yasymm\Big)\, \oplus\,N_\symm~\big(\Ysymm\big)~,
\label{Rrep}
\end{align} 
for which the interaction action can be systematically rewritten in terms of fundamental traces using the results of \cite{Billo:2019fbi,Beccaria:2020hgy}.
This reformulation renders the resulting matrix model amenable to a controlled large-$N$ analysis.

For the representation (\ref{Rrep}), the $\beta$-function coefficient (\ref{b0def}) becomes
\begin{align}
    b_0=2N(1-N_\adj)-N_\fund-(N-2)\,N_\asymm-(N+2)\,N_\symm~.
    \label{b0R}
\end{align}
The choice $N_\adj=1$, $N_\fund=N_\asymm=N_\symm=0$ corresponds to $\mathcal{N}=4$ SYM, while genuinely $\mathcal{N}=2$ asymptotically free theories are characterized by $N_\adj=0$. In this work, we focus primarily on the class of theories defined by
\begin{align}
    N_\adj=N_\symm=0~,\quad N_\asymm=2~,\quad 0\leq N_\fund\leq4 ~,
    \label{Dhat}
\end{align}
which we refer to as the $\widehat{\mathbf{D}}$-theory. For this class, the $\beta$-function coefficient simplifies to 
\begin{align}
    b_0=4-N_\fund\ .
    \label{b0Dhat}
\end{align}
When $N_\fund=4$, the theory becomes superconformal and coincides with the so-called \textbf{D}-theory, previously studied in \cite{Beccaria:2021ism,Beccaria:2022ypy} and more recently in the context of integrated correlators in \cite{Billo:2024ftq,DeLillo:2025eqg,DeLillo:2025stg}.
Another superconformal model in this class is obtained for
\begin{align}
 N_\adj=N_\fund=0~,\quad N_\asymm=N_\symm=1~,
 \label{Etheory}
\end{align}
corresponding to the so-called \textbf{E}-theory, extensively investigated in
\cite{Beccaria:2020hgy,Beccaria:2021vuc,Beccaria:2021hvt,Billo:2022xas,Beccaria:2022ypy,Billo:2022gmq,Billo:2022fnb,Billo:2023kak,Beccaria:2023kbl}.

The \textbf{D}- and \textbf{E}-theories are the only $\mathcal{N}=2$ superconformal models in this family whose number of hypermultiplets is independent of $N$. Moreover, the $\widehat{\mathbf{D}}$-theory with $0\leq N_\fund<4$ is the unique asymptotically free $\mathcal{N}=2$ model in this class whose $\beta$-function coefficient remains $N$-independent.

\subsection{The interaction action}
When the matter representation $\mathcal{R}$ is of the form \eqref{Rrep} the primed traces appearing in \eqref{SintR} can be expressed in terms of fundamental traces as \cite{Billo:2019fbi,Beccaria:2020hgy}
\begin{align}
    \label{eq:primed traces in terms of fundamental}
    \Tr_{\mathcal{R}}^{\prime} a^{2k}
    &\equiv \Tr_{\mathcal{R}} a^{2k} - \Tr_{\adj} a^{2k} \notag\\
    &= \frac{1}{2}\sum_{\ell=2}^{2k-2}
    \binom{2k}{\ell}
    \Big[N_{\asymm}+N_{\symm}
    + 2(-1)^\ell (N_{\adj}-1)\Big]
    \tr a^{\ell}\,\tr a^{2k-\ell}
    \notag\\[1mm]
    &\quad
    + \Big[(2^{2k-1}-2)(N_{\symm}-N_{\asymm}) - b_0\Big]\,\tr a^{2k} ~,
\end{align}
with $b_0$ given in \eqref{b0R}. 
For $\mathcal{N}=4$ SYM this expression vanishes identically, reflecting the purely Gaussian nature of the corresponding matrix model.
For the $\widehat{\mathbf{D}}$-theory, we tune the coefficients $N_\symm$ and $N_\asymm$ according to \eqref{Dhat} and obtain
\begin{align}
    \label{traRDhat}
    \Tr_{\mathcal{R}}^{\prime} a^{2k}
    = 2\sum_{\ell=1}^{k-2}
    \binom{2k}{2\ell+1}\,
    \tr a^{2\ell+1}\,\tr a^{2k-2\ell-1}
    - \big(2^{2k}-4+b_0\big)\,\tr a^{2k}~.
\end{align} 
In this expression the double-trace terms coincide with those of the \textbf{E}-theory, as can be verified by setting 
$N\symm=N\asymm=1$ in (\ref{eq:primed traces in terms of fundamental}). Moreover, (\ref{traRDhat}) correctly reproduces the superconformal \textbf{D}-theory when $b_0=0$. Substituting  (\ref{traRDhat}) in (\ref{SintR}), the interaction action of the $\widehat{\mathbf{D}}$-theory can be written as the sum of two distinct contributions
\begin{align}
    \label{Sdhat}
    S_{\widehat{\mathbf{D}}}
    = S_{{\mathbf{E}}} + S_{\text{s-t}}\ .
\end{align}
Introducing the rescaled operators
\begin{align}
    \cO_k = \tr \Big(\frac{a}{\sqrt{N}}\Big)^k~,
    \label{Ok}
\end{align}
one finds that $S_{\mathbf{E}} $ and  $S_{\text{s-t}}$ are given by
\begin{align}
    \label{SEis}
     S_{\mathbf{E}} &=  -\frac{1}{2} \sum_{j_1,j_2=1}^{\infty} 
C_{j_1j_2}\,\cO_{2j_1+1}\,\cO_{2j_2+1} ~,\\[2mm]
S_{\text{s-t}}
     &= - \sum_{i=1}^{\infty} J_i\,\cO_{2i}~,\label{defSst}
\end{align}
where the semi-infinite matrix $C_{j_1j_2}$ is \cite{Beccaria:2023kbl}
\begin{align}
    \label{eq:C^-}
    C_{j_1j_2}= 8(-1)^{j_1+j_2+1}
    \Big(\frac{\lambda}{8\pi^2}\Big)^{j_1+j_2+1}
    \frac{\Gamma(2j_1+2j_2+2)}{\Gamma(2j_1+2)\Gamma(2j_2+2)}
    \zeta(2j_1+2j_2+1)~,
\end{align}
while the sources $J_i$ are
\begin{align}
\label{eq:source K}
J_1=0\quad\mbox{and}\quad
J_i= (-1)^{i+1}\Big(\frac{\lambda}{8\pi^2}\Big)^{i}\,\,\frac{\zeta(2i-1)}{i}
\big(2^{2i}-4+b_0\big) \quad \mbox{for}~i\geq 2~.
\end{align}

\subsection{The partition function}
Using (\ref{SEis}) and (\ref{defSst}) in (\ref{Zresc}), the partition function of the $\widehat{\mathbf{D}}$-theory can be recast as
\begin{align}
	\label{eq: Z dhat}
\cZ_\Dhat
=\cZ_\NN \int \!\mathcal{D}a ~\rme^{-\tr a^2 +\frac{1}{2}
\sum\limits_{j_1\!,j_2=1}^\infty \!
C_{j_1j_2}\,\cO_{2j_1+1}\,\cO_{2j_2+1}+
\sum\limits_{i=1}^\infty \!
J_i \,\cO_{2i} }~.
\end{align}
Following \cite{Beccaria:2023kbl}, we linearize the quadratic term in \eqref{eq: Z dhat} via a Hubbard--Stratonovich transformation. Introducing auxiliary sources $K_j$ coupled to the odd operators leads to 
\begin{align}
	\label{eq:Z dhat cumulant}
	\cZ_\Dhat
	&= \frac{\cZ_\NN}{\sqrt{\det C}}
	\int\! \cD K ~
	\rme^{-\frac{1}{2}\sum\limits_{j_1,j_2=1}^\infty (C^{-1})_{j_1j_2}\,K_{j_1}K_{j_2}} 
    \int\! \mathcal{D}a~
    \rme^{-\tr a^2
    +\sum\limits_{j=1}^\infty K_j\,\cO_{2j+1}
    +\sum\limits_{i=1}^\infty J_i \,\cO_{2i}}~.
\end{align}
In the Gaussian integral over $a$, the sources $K_j$ and $J_i$ enter on the same footing. It is therefore natural to introduce collective sources $\rho_k$ defined by
\begin{equation}
\rho_{2i} \equiv J_i ~, \qquad
\rho_{2j+1} \equiv K_j ~,
\label{JK}
\end{equation}
so that the matrix integral in (\ref{eq:Z dhat cumulant}) can be expressed as a cumulant expansion
\begin{align}
    \label{cumexp}
 Z_0 =   \int\!\mathcal{D}a~
    \rme^{-\tr a^2 + \rho_k \cO_k}
    =\exp\bigg[\sum\limits_{L=1}^\infty\frac{1}{L!}\,\rho_{k_1}\ldots \rho_{k_L}\,G_{k_1\ldots k_L}
   \bigg]~.
\end{align}
Here, repeated indices are summed from 2 to $\infty$, and 
\begin{align}
    \label{defGcorr}
    G_{k_1\ldots k_L}
    = \big\langle \cO_{k_1}\cdots \cO_{k_L} \big\rangle_0^{c}~,
\end{align}
denotes the connected Gaussian correlator of $L$ single-trace operators. Owing to the symmetry of the Gaussian measure under $a\to -a$, 
these correlators vanish unless the number of odd operators is even. Writing $L=n+2m$, where $n$ is the number of even operators and $2m$ the number of odd ones, we introduce the notation
\begin{equation}
    \label{defQs}    G_{2i_1\ldots 2i_n;2j_1+1 \ldots 2j_{2m}+1}\,\equiv\,Q^{(n|2m)}_{i_1\ldots i_n;j_1\ldots j_{2m}}~,
\end{equation}
with $Q^{(0|0)}=0$. When indices are not necessary, we will simply write $Q^{(n|2m)}$.

Separating the sources $\rho_k$ into $J_i$ and $K_j$ according to the index parity, we find
\begin{align}
Z_0 &= \exp \left[    \sum_{m=0}^\infty\frac{1}{(2m)!}\,K_{j_1}\ldots K_{j_{2m}}\,\widehat{Q}^{(2m)}_{j_1\ldots j_{2m}}\right]\ ,   \label{cumexp1} 
\end{align}
where
\begin{align}
     \widehat{Q}^{(2m)}_{j_1,\ldots,j_{2m}}&=\sum_{n=0}^\infty \frac{1}{n!}\,J_{i_1}\ldots J_{i_n}\,Q^{(n|2m)}_{i_1\ldots i_n;j_1\ldots j_{2m}} \ ,
    \label{Qhat}
\end{align}
and repeated indices are summed from 1 to $\infty$. Using these expressions, the partition function (\ref{eq:Z dhat cumulant}) becomes 
\begin{align}
	\label{Z dhat1}
	\cZ_\Dhat
	&= \frac{\cZ_\NN}{\sqrt{\det C}}~\rme^{\widehat{Q}^{(0)}}\!\!
	\int\! \cD K ~
	\rme^{-\frac{1}{2}\, \left(C^{-1}-\widehat{Q}^{(2)}\right)_{j_1j_2}
    \,K_{j_1}K_{j_2}+\sum\limits_{m=2}^{\infty}\frac{1}{(2m)!}\,K_{j_1}\ldots K_{j_{2m}}\,\widehat{Q}^{(2m)}_{j_1,\ldots,j_{2m}}} \ .
\end{align}
Up to overall factors, the $\widehat{\mathbf{D}}$-theory partition function takes the form of an effective theory for the variables
$K_j$, with propagator
\begin{equation}
    \label{defXprop}
        X \,\equiv\,\big(C^{-1}-\widehat{Q}^{(2)}\big)^{-1}=
        \big(1- C\,\widehat{Q}^{(2)}\big)^{-1} C~,
\end{equation}
and interaction vertices $\widehat{Q}^{(2m)}$. According to their definition (\ref{Qhat}), these vertices depend on the Gaussian correlators $Q^{(n|2m)}$ given in (\ref{defQs}) which, in the large-$N$ limit, admit the following topological expansion \cite{Beccaria:2023kbl}
\begin{align}
Q^{(n|2m)}_{i_1,\ldots,i_n;j_1,\ldots,j_{2m}}&=\frac{\beta_{i_1}\ldots\beta_{i_n}\gamma_{j_1}\ldots\gamma_{j_{2m}}}{N^{n+2m}}\,
\sum_{g=0}^\infty N^{2-2g}\,P^{(n|2m)}_{g}(i_1,\ldots,i_n;j_1,\ldots,j_{2m})~.
\label{expansionmain}
\end{align}
Here $P_g^{(n|2m)}$ are symmetric polynomials in $\{i_1,\ldots,i_n\}$ and $\{j_1,\ldots,j_{2m}\}$ of degree
$n+2m+3g-3$ (see Appendix \ref{sec:GC} for details), while the coefficients $\beta_i$ and $\gamma_i$ are
\begin{align}
\beta_i&=\frac{1}{2^{i-1}}\frac{\Gamma(2i)}{\Gamma(i)^2}~,
    \label{betai}\\[2mm]
    \gamma_j&=\frac{1}{2^{j+\frac{1}{2}}}\frac{\Gamma(2j+2)}{\Gamma(j+2)\Gamma(j)}~.
    \label{gammaj}
\end{align}
From (\ref{expansionmain}) it follows that the effective vertices $\widehat{Q}^{(2m)}$ with $m\ge1$ scale as $N^{2-2m}$, while $\widehat{Q}^{(0)}$ scales 
as $N$. As a result, at any fixed order in the $1/N$-expansion, only finitely many vertices can contribute, ensuring that the large-$N$ analysis is well-controlled.
As we will shortly show, reformulating the matrix model of the $\widehat{\mathbf{D}}$-theory in terms of the effective theory (\ref{Z dhat1})
provides an efficient framework for deriving explicit and exact expressions for a variety of observables in the large-$N$ limit, as well as for analyzing their large-$\lambda$ regime. 

Finally, we observe that the dependence of the partition function on the external sources $J_i$ is through both $X$ and $\widehat{Q}^{(2m)}$. Setting $J_i=0$ in (\ref{Z dhat1}) switches off the single-trace interactions and yields the matrix model for the \textbf{E}-theory. In this limit
\begin{equation}
    \label{QttoQ}
        \widehat{Q}^{(2m)}\big|_{J = 0} = Q^{(0|2m)}
\end{equation}
and the partition function reduces to \cite{Beccaria:2023kbl}
\begin{equation}
    \label{ZeffJE}
        \cZ_\E =  
        \frac{\cZ_\NN}{\sqrt{\det C}} \int \!\mathcal{D}K~\rme^{- \frac 12 \big(C^{-1} - Q^{(0|2)}\big)_{j_1 j_2}\, K_{j_1} K_{j_2}
        + \sum\limits_{m=2}^\infty \frac{1}{(2m)!}\,K_{j_1}\ldots  K_{j_{2m}} \,Q^{(0|2m)}_{j_1\ldots j_{2m}} }~.	
\end{equation}

\section{The free energy and the Wilson loop in the \texorpdfstring{$\widehat{\mathbf{D}}$}{}-theory at large \texorpdfstring{$N$}{}}
\label{secn:freeenergy+WL}
In this section, we introduce the matrix formulation of the two observables of interest: the free energy and the expectation value of the $\frac{1}{2}$-BPS Wilson loop. 

\subsection{The free energy}
\label{subsec:freeenergy}
The free energy of the $\widehat{\mathbf{D}}$-theory on $S^4$ is among the simplest quantities we can consider and is defined by
\begin{equation}
    \label{eq:FD}
        F_\Dhat=-\log\cZ_\Dhat~.
\end{equation}
Starting from the effective formulation \eqref{Z dhat1}, we observe that $F_\Dhat$ receives four distinct contributions: (\emph{i}) the free energy of $\mathcal{N}=4$ SYM 
\begin{align}
    F_\NN=-N^2\,\frac{\log\lambda}{2}+\frac{\log\lambda}{2}~,
    \label{FN=4}
\end{align}
which follows from the Gaussian matrix integral (see for example \cite{Russo:2012ay})\,\footnote{As noted in Appendix~A of \cite{Beccaria:2021ism}, this localization result holds in a specific subtraction scheme where the logarithmic UV divergence of the free energy is removed.}; (\emph{ii}) a term proportional to $\widehat{Q}^{(0)}$; (\emph{iii}) a contribution arising from the determinant of the quadratic kernel $\big(C^{-1}-\widehat{Q}^{(2)}\big)$; and (\emph{iv}) a sum of connected vacuum diagrams built from the interaction vertices $\widehat{Q}^{(2m)}$.  
Graphically, we represent these vertices as blobs with $2m$ legs, while contractions with the propagator $X$ are depicted by continuous lines. In this notation, the free energy takes the form
\begin{align}
    \label{eq:FD1}
        F_\Dhat&= 
        F_\NN -\widehat{Q}^{(0)}+\frac{1}{2}\tr \log\big(1- C\,\widehat{Q}^{(2)}\big)\notag\\ 
        &\quad+~
        \begin{tikzpicture}[scale=0.8,baseline=-0.65ex]
        \begin{feynman}
        	\vertex (O) at (0,1); 
        	\vertex (OO) at (0,0);
        	\vertex (OOO) at (0,-1);			
        	\diagram*{
        		(OO)--[plain, half left,looseness=1.1] (OOO),
        		(OO)--[plain, half right,looseness=1.1] (OOO),
        			(OO)--[plain, half left,looseness=1.1] (O),
        		(OO)--[plain, half right,looseness=1.1] (O),
        	};
        \end{feynman}
        \filldraw[fill=blue!20, draw=black] (0,0) circle (0.4cm);
        \begin{feynman}
				\vertex (o) at (0,0) {\scalebox{.65}{$\widehat{Q}^{(4)}$}};
			\end{feynman}
        \end{tikzpicture}~~+
\mathord{
	\begin{tikzpicture}[scale=0.7,baseline=-0.65ex]
		\begin{feynman}
			\vertex (O) at (-1.3,0.8); 
			\vertex (A) at (1.3,0.8);
			\vertex (OO) at (0,0);
			\vertex (OOO) at (0,-1.4);			
			\diagram*{
				(OO)--[plain, half left,looseness=0.9] (OOO),
				(OO)--[plain, half right,looseness=0.9] (OOO),
				(OO)--[plain, half left,looseness=0.9] (O),
				(OO)--[plain, half right, looseness=0.9] (O),
				(OO)--[plain, half left,looseness=0.9] (A),
				(OO)--[plain, half right, looseness=0.9] (A),
			};
		\end{feynman}
		\filldraw[fill=blue!20, draw=black] (0,0) circle (0.45cm);
        \begin{feynman}
				\vertex (o) at (0,0) {\scalebox{.65}{$\widehat{Q}^{(6)}$}};
			\end{feynman}
\end{tikzpicture} }~~ +~~  \mathord{
\begin{tikzpicture}[scale=0.8,baseline=-0.65ex]
\begin{feynman}
	\vertex (O) at (-1,1); 
	\vertex (B) at (1.2,0);
	\vertex (A) at (2.4,0);
	\vertex (OO) at (0,0);
	\vertex (OOO) at (-1.2,0);			
	\diagram*{
		(OO)--[plain, half left] (OOO),
		(OO)--[plain, half right] (OOO),
		(OO)--[plain, half left] (B),
		(OO)--[plain, half right] (B),
		(B)--[plain, half left] (A),
		(B)--[plain, half right ] (A),
	};
\end{feynman}
\filldraw[fill=blue!20, draw=black] (0,0) circle (0.4cm);
\filldraw[fill=blue!20, draw=black] (1.2,0) circle (0.4cm);
\begin{feynman}
				\vertex (o) at (0,0) {\scalebox{.65}{$\widehat{Q}^{(4)}$}};
                \vertex (oo) at (1.2,0) {\scalebox{.65}{$\widehat{Q}^{(4)}$}};
			\end{feynman}
\end{tikzpicture} }~~+~\ldots~,
\end{align} 
According to (\ref{QttoQ}), replacing $\widehat{Q}^{(2m)}$ with $Q^{(0|2m)}$ yields the free energy of the \textbf{E}-theory.
The first few orders in the large-$N$ expansion of this quantity were computed in \cite{Beccaria:2021vuc,Beccaria:2022ypy,Beccaria:2023kbl}.
Here we revisit and extend these calculations in a systematic way, preparing the ground for the strong-coupling analysis presented in Section~\ref{secn:free energy}.
 
\paragraph{The function $\widehat{Q}^{(0)}$:}
The second contribution in (\ref{eq:FD1}) is $-\widehat{Q}^{(0)}$. Using \eqref{Qhat} we find
\begin{align}
    -\widehat{Q}^{(0)}=-\sum_{n=1}^\infty \frac{1}{n!}\,J_{i_1}\ldots J_{i_n}\,Q^{(n|0)}_{i_1\ldots i_n}~,
    \label{Qhat0}
\end{align} 
where the coefficients $Q^{(n|0)}_{i_1\ldots i_n}$ are the Gaussian correlators defined in (\ref{defGcorr}).
From (\ref{expansionmain}) it follows that 
\begin{align}
    Q^{(n|0)}_{i_1 \ldots i_n}=\frac{\beta_{i_1}\ldots \beta_{i_n}}{N^n}\,\sum_{g=0}^\infty N^{2-2g}\,P^{(n|0)}_g(i_1,\ldots,i_n)
    \label{Qn0exp}
\end{align}
where $\beta_i$ is given in (\ref{betai}) and $P^{(n|0)}_g(i_1,\ldots,i_n)$ is a symmetric polynomial of degree $n+3g-3$ in its arguments. Substituting (\ref{Qn0exp}) into (\ref{Qhat0}), we obtain
\begin{align}
    -\widehat{Q}^{(0)}&=-\sum_{g=0}^\infty\sum_{n=1}^\infty \frac{N^{2-2g-n}}{n!} \prod_{\ell=1}^n \big(\beta_{i_\ell}\,J_{i_\ell}\big) P^{(n|0)}_g(i_1,\ldots,i_n)\notag\\[2mm]
    &=N\,q_1+q_2+\frac{1}{N}\,q_3+\frac{1}{N^2}\,q_4
    +\ldots\label{Qhat0exp1}\ ,
\end{align}
where, in the first line, there is an implicit sum over the indices $i_\ell\geq 2$. The general structure of the expansion coefficients $q_\ell$ is
\begin{equation}
    \label{eq:qellapp}
    q_\ell = -\sum_{g=0}^{\lfloor\frac{\ell-1}{2}\rfloor}\frac{1}{(\ell-2g) !}\prod_{q=1}^{\ell-2g}\left(\beta_{i_q} J_{i_q}\right) P_g^{(\ell-2g|0)}(i_1,\ldots,i_{\ell-2g}) \ .
\end{equation} 
For the first few values of $\ell$, one readily finds
\begin{subequations}
\begin{align}
    q_1&=-\big(\beta_i J_i \big)P^{(1|0)}_0(i)~,\label{q1expl}\\[1mm]
    q_2&=-\frac{1}{2}\big(\beta_{i_1} J_{i_1}\big)\big( \beta_{i_2} J_{i_2}\big) P^{(2|0)}_0(i_1,i_2)~,\label{q2}
    \\[1mm]
    q_3&=-\frac{1}{3!}\big(\beta_{i_1} J_{i_1} \big)\big(\beta_{i_2} J_{i_2}\big)\big(\beta_{i_3} J_{i_3}\big) P^{(3|0)}_0(i_1,i_2,i_3)-\big(\beta_i J_i \big)P^{(1|0)}_1(i)~.\label{q3}
\end{align}
\label{qi}%
\end{subequations}
These coefficients depend on the 't Hooft coupling through the sources $J_i$ according to (\ref{eq:source K}). 
For instance, using the definition (\ref{betai}) for $\beta_i$ and the explicit expression for the polynomial $P^{(1|0)}_0$ given in (\ref{P0n0}), 
we obtain
\begin{align}
\label{q1explicit}
    q_1=-\sum_{i=2}^\infty\frac{\beta_i J_i }{i(i+1)}=2\sum_{i=2}^\infty(-1)^{i}\Big(\frac{\lambda}{16\pi^2}\Big)^{i}\,\frac{\Gamma(2i)\,\zeta(2i-1)}{\Gamma(i+1)\Gamma(i+2)}
\big(4^{i}-4+b_0\big)\ .
\end{align}
To streamline higher-order expressions, we introduce the one-parameter family of functions
\begin{align}\label{Yell}
		\mathcal{Y}_{a}
        &= \sum_{i=2}^{\infty}  \beta_i J_i \,i^{a-1}
        =-2\sum_{i=2}^{\infty}(-1)^i\Big(\frac{\lambda}{16\pi^2}\Big)^{\!i}\,i^{a}\,\frac{\Gamma(2i)\,\zeta({2i-1})}{\Gamma^2(i+1)} \,\big(4^{i}-4+b_0 \big)~.
\end{align} 
Exploiting the explicit form of the polynomials $P_g^{(n|0)}$ reported in Appendix~\ref{sec:GC}, one can show that
\begin{align}
    \partial_\lambda q_2=-\frac{\mathcal{Y}_1^{\,2}}{2\lambda}~,\quad
    q_3=-\frac{\mathcal{Y}_1^{\,3}}{6}-\frac{\mathcal{Y}_2}{12}+\frac{7\mathcal{Y}_1}{12}~.
    \label{q2q3}
\end{align}
Similar expressions can be easily derived for $q_\ell$ with $\ell>3$.

\paragraph{Trace-log and vacuum diagrams:}
Let us turn our attention to the trace-log term in (\ref{eq:FD1}) which admits the expansion
\begin{equation}
    \label{trlogexpanded}
        \frac{1}{2}\tr \log\big(1- C\,\widehat{Q}^{(2)}\big)= -\sum_{k=1}^\infty \frac{1}{2k}\tr \big(C\,\widehat{Q}^{(2)}\big)^k 
\end{equation}
where
\begin{equation}
    \label{trCQk}
        \tr \big(C\,\widehat{Q}^{(2)}\big)^k =
        \prod_{p=0}^{k-1} C_{j_{2p}\,j_{2p+1}}\,\widehat{Q}^{(2)}_{j_{2p+1}\,j_{2p+2}}~.
\end{equation}
Here all repeated indices are summed and the identification $j_{2k}\equiv j_0$ is enforced.
The kernel $\widehat{Q}^{(2)}$ inherits a large-$N$ expansion from connected correlators (\ref{expansionmain}), namely
\begin{align}
   \widehat{Q}^{(2)}_{j_1,j_2}&= \sum_{n=0}^\infty\frac{1}{n!}J_{i_1}\ldots J_{i_n}\,Q^{(n|2)}_{i_1\ldots i_n;j_1j_2}\notag\\[2mm]
   &=\sum_{g=0}^\infty\sum_{n=0}^\infty \frac{N^{-2g-n}}{n!} \prod_{\ell=1}^n \big(\beta_{i_\ell}\,J_{i_\ell}\big)
   P^{(n|2)}_g(i_1,\ldots,i_n;j_1,j_2)\,\gamma_{j_1}\gamma_{j_2}~,
\end{align}
where $\gamma_i$ is given in (\ref{gammaj}) and $P_g^{(n|2)}$ is a polynomial of degree $n+3g-1$, symmetric in $\{i_1\ldots i_n\}$ and in $\{j_1,j_2\}$. Exploiting the explicit expression for these polynomials (see Appendix~\ref{sec:GC} for a few examples) and following the approach developed in \cite{Beccaria:2023kbl} for the \textbf{E}-theory, we find
\begin{align}
        \frac{1}{2}\tr \log\big(1- C\,\widehat{Q}^{(2)}\big) &= \frac{1}{2}\tr \log\big(1- C\,{Q}^{(0|2)}\big)-\frac{1}{N}\Big(\frac{\mathcal{Y}_1}{8}\,w_{0,0}\Big)\notag\\[2mm]
        &\quad-\frac{1}{N^2}\Big(\frac{\mathcal{Y}_1\,\mathcal{Y}_2}{8}w_{0,0}+\frac{\mathcal{Y}_1^{\,2}}{8}w_{0,1}+\frac{\mathcal{Y}_1^{\,2}}{64}w_{0,0}^2\Big)+\ldots~.
        \label{trloglargeN}
\end{align}
Here $w_{n,m}$ are the quantities introduced in \cite{Beccaria:2023kbl} to describe the matrix elements of the resolvent of the Bessel operator related to the \textbf{E}-theory. Furthermore, the trace-log term in the right-hand side of (\ref{trloglargeN}) is proportional to the Fredholm determinant of this Bessel operator and admits a large-$N$ expansion in powers of $1/N^2$, as shown in Sections 3 and 4 of
\cite{Beccaria:2023kbl}.

The vacuum diagrams in (\ref{eq:FD1}) can be studied in the same way. For example one finds
\begin{align}
    \begin{tikzpicture}[scale=0.8,baseline=-0.65ex]
        \begin{feynman}
        	\vertex (O) at (0,1); 
        	\vertex (OO) at (0,0);
        	\vertex (OOO) at (0,-1);			
        	\diagram*{
        		(OO)--[plain, half left,looseness=1.1] (OOO),
        		(OO)--[plain, half right,looseness=1.1] (OOO),
        			(OO)--[plain, half left,looseness=1.1] (O),
        		(OO)--[plain, half right,looseness=1.1] (O),
        	};
        \end{feynman}
        \filldraw[fill=blue!20, draw=black] (0,0) circle (0.4cm);
        \begin{feynman}
				\vertex (o) at (0,0) {\scalebox{.65}{$\widehat{Q}^{(4)}$}};
			\end{feynman}
        \end{tikzpicture}=-\frac{1}{N^2}\Big(\frac{1}{32}w_{0,0}^2+\frac{1}{32}w_{0,0}\,w_{0,1}\Big)+\ldots~.
        \label{Q4diagram}
\end{align}
Diagrams containing vertices $\widehat{Q}^{(2m)}$ with $m>2$ contribute only at increasingly subleading orders in the large-$N$ expansion, and will not
be needed in what follows.

\subsection{The circular Wilson loop}
\label{subsec:cwl}
The second observable we consider is the expectation value of the $\frac{1}{2}$-BPS Wilson loop
\begin{align}
    \label{eq:WD}
        W= \frac{1}{N} \tr \mathcal{P}\exp \Bigg\{ g\oint_C \!ds\,\bigg[\ii A_\mu(x)\,\dot{x}^\mu(s)+\frac{R}{\sqrt{2}}\,\left(\phi(x)+\bar{\phi}(x)\right)\bigg] \Bigg\}
\end{align}
where the gauge field $A_\mu$, and the vector-multiplet scalars $\phi$ and $\bar{\phi}$ are integrated along a circle $C$ of radius $R$, parametrized by $s\in[0,2\pi)$ with $x(s)^2=\dot{x}^2(s)=R^2$.

Using supersymmetric localization \cite{Pestun:2007rz}, the Wilson loop (\ref{eq:WD}) admits the matrix-model representation
$\frac{1}{N} \tr \rme^{2\pi R a}$, which after the rescaling (\ref{rescaling}) becomes
\begin{align}
   \mathcal{W}= \frac{1}{N} \tr\rme^{\sqrt{\frac{\lambda}{2N}}\,a}=\frac{1}{N} \sum_{p=0}^\infty\frac{1}{p!}\,\Big(\frac{\lambda}{2N}\Big)^{\frac{p}{2}} \tr a^p~.
     \label{Wmm}
\end{align}
Taking the vacuum expectation value in the $\widehat{\mathbf{D}}$-theory, we obtain
\begin{align}
    \big\langle \mathcal{W} \big\rangle_\Dhat=1+\frac{1}{N}\sum_{i=1}^\infty\frac{1}{(2i)!}\,\Big(\frac{\lambda}{2}\Big)^i\,\big\langle \mathcal{O}_{2i}\big\rangle_\Dhat
    \label{WtoG}
\end{align}
where the operators $\mathcal{O}_i$ are defined in (\ref{Ok}) and we used the vanishing of odd-trace expectation values. 
The evaluation of $\big\langle \mathcal{W} \big\rangle_\Dhat$ therefore reduces to computing the one-point functions of the even operators $\mathcal{O}_{2i}$. From (\ref{eq: Z dhat}), these are obtained as
\begin{equation}
    \label{OtoDJ}
       \big\langle \mathcal{O}_{2i}\big\rangle_\Dhat = \frac{\partial}{\partial J_i} \log \cZ_{\widehat{\mathbf{D}}}\ .
\end{equation}
Using the effective theory (\ref{Z dhat1}) for the sources $K_j$, and recalling that the sources $J_i$ enter only through the interaction vertices $\widehat{Q}^{(2m)}$, we find
\begin{equation}
    \label{cO2neff}
        \big\langle \mathcal{O}_{2i}\big\rangle_\Dhat = \sum_{m=0}^\infty \frac{1}{(2m)!}\, \widehat{R}^{(2m)}_{i;j_1, \ldots, j_{2m}} ~\big\langle\!\!\big\langle K_{j_1}\ldots K_{j_{2m}}\big\rangle\!\!\big\rangle
\end{equation}
where 
\begin{align}
    \label{defRtilde}
    \widehat{R}^{(2m)}_{i;j_1, \ldots, j_{2m}} \,\equiv\, \frac{\partial }{\partial J_i}\widehat{Q}^{(2m)}_{j_1 \ldots j_{2m}}
    =\sum_{p=0}^\infty \frac{1}{p!} \,J_{k_1}\ldots J_{k_p}\,Q^{(p+1|2m)}_{i,k_1,\ldots, k_p;j_1,\ldots,j_{2m}}~,
\end{align}
and the double brackets $\big\langle\!\!\big\langle ~\big\rangle\!\!\big\rangle$ denote expectation values in the effective theory (\ref{Z dhat1}). 

The correlators $\big\langle\!\!\big\langle K_{j_1}\ldots K_{j_{2m}}\big\rangle\!\!\big\rangle$ appearing in (\ref{cO2neff}) can be computed by standard Feynman diagram techniques using the propagator $X$ and the vertices $\widehat{Q}^{(2m)}$, defined respectively in (\ref{defXprop}) and (\ref{Qhat}).
As an illustration, we can consider the two-point correlator that admits the following diagrammatic expansion
\begin{align}
    \label{JJ}
		\big\langle\!\!\big\langle K_{j_1} K_{j_2} \big\rangle\!\!\big\rangle
        &=\mathord{
			\begin{tikzpicture}[scale=0.7, baseline=-0.65ex]
				\begin{feynman}
					\vertex (A) at (-1.,0) ;
					\vertex (C) at (1.,0) ;
					\vertex (V) at (1,0.3) {\footnotesize{$j_2$}};
					\vertex (V) at (-1,0.3) {\footnotesize{$j_1$}};
					\diagram*{
						(A) -- [plain] (C),
					};
				\end{feynman}
			\end{tikzpicture} 
		}  ~+~	\mathord{
		\begin{tikzpicture}[scale=0.8,baseline=-0.65ex]
			\begin{feynman}
				\vertex (O) at (0,0);
                \vertex (OO) at (0,1);
				\vertex (A) at (0,0); 
				\vertex (AA) at (-1,1);
				\vertex (AAA) at (1,1);
				\vertex (B) at (-1.,0) ;
				\vertex (C) at (1.,0) ;
				\vertex (V) at (1,0.3) {\footnotesize{$j_2$}};
				\vertex (V) at (-1,0.3) {\footnotesize{$j_1$}};
				\diagram*{
					(O) -- [plain, half left] (OO),
                    (O) -- [plain, half right ] (OO),
						(B) -- [plain] (C),
				};
			\end{feynman}
			\filldraw[fill=blue!20, draw=black] (0,0) circle (0.4cm);
            \begin{feynman}
				\vertex (o) at (0,0) {\scalebox{.65}{$\widehat{Q}^{(4)}$}};
			\end{feynman}
	\end{tikzpicture} }~+~ \mathord{
		\begin{tikzpicture}[scale=0.8,baseline=-0.65ex]
			\begin{feynman}
				\vertex (O) at (0,0);
                \vertex (OO) at (0,1);
                 \vertex (OOO) at (0,-1);
				\vertex (A) at (0,0); 
				\vertex (AA) at (-1,1);
				\vertex (AAA) at (1,1);
				\vertex (B) at (-1.,0) ;
				\vertex (C) at (1.,0) ;
				\vertex (V) at (1,0.3) {\footnotesize{$j_2$}};
				\vertex (V) at (-1,0.3) {\footnotesize{$j_1$}};
				\diagram*{
					(O) -- [plain, half left] (OO),
                    (O) -- [plain, half right ] (OO),
						(B) -- [plain] (C),
                        (O) -- [plain, half left] (OOO),
                    (O) -- [plain, half right ] (OOO),
				};
			\end{feynman}
			\filldraw[fill=blue!20, draw=black] (0,0) circle (0.4cm);
            \begin{feynman}
				\vertex (o) at (0,0) {\scalebox{.65}{$\widehat{Q}^{(6)}$}};
			\end{feynman}
	\end{tikzpicture} }~+~\ldots~ \notag \\[2mm]
    &= X_{j_1 j_2} + \frac{1}{2} \widehat{Q}^{(4)}_{mnpq}\,X_{j_1 m}X_{np}X_{qj_2}+\frac{1}{24}  \widehat{Q}^{(6)}_{mnpq rs}\,X_{j_1 m}X_{np}X_{qr} X_{s j_2}+\ldots 
\end{align} 
where the numerical coefficients are the symmetry factors of the corresponding diagrams. Higher-point correlators are obtained analogously. 

According to (\ref{cO2neff}), the computation of $\big\langle \mathcal{O}_{2i}\big\rangle_\Dhat$ amounts to contracting the external legs of  $\big\langle\!\!\big\langle K_{j_1}\ldots K_{j_{2m}}\big\rangle\!\!\big\rangle$ with the coefficients $\widehat{R}^{(2m)}_{i;j_1, \ldots, j_{2m}}$. We graphically represent these coefficients as blobs with $2m + 1$ legs, one of which (associated to the index $i$) is depicted by a dashed line. This leads to the following diagrammatic representation of $\big\langle \mathcal{O}_{2i}\big\rangle_\Dhat$ 
\begin{equation}
    \label{O2ndiag}
       \big\langle \mathcal{O}_{2i}\big\rangle_\Dhat= 
         \mathord{
		\begin{tikzpicture}[scale=0.8,baseline=-0.65ex]
			\begin{feynman}
				\vertex (O) at (0,0);
				\vertex (AA) at (-1,1);
				\vertex (AAA) at (1,1);
				\vertex (B) at (-1.,0) ;
                \vertex (D) at (-0.9,0.3)  {\footnotesize{$i$}};
				\diagram*{
						(B) -- [thick,dashed] (O),
				};
			\end{feynman}
			\filldraw[fill=red!20, draw=black] (0,0) circle (0.4cm);
            \begin{feynman}
				\vertex (o) at (0,0) {\scalebox{.65}{$\widehat{R}^{(0)}$}};
			\end{feynman}
	\end{tikzpicture} }       
        \,+\, 	\mathord{
		\begin{tikzpicture}[scale=0.8,baseline=-0.65ex]
			\begin{feynman}
				\vertex (O) at (0,0);
                \vertex (OO) at (1.2,0);
				\vertex (A) at (0,0); 
				\vertex (AA) at (-1,1);
				\vertex (AAA) at (1,1);
				\vertex (B) at (-1.,0) ;
                \vertex (D) at (-0.9,0.3)  {\footnotesize{$i$}};
				\vertex (C) at (1.,0) ;
				\diagram*{
					(O) -- [plain, half left] (OO),
                    (O) -- [plain, half right ] (OO),
						(B) -- [thick,dashed] (O),
				};
			\end{feynman}
			\filldraw[fill=red!20, draw=black] (0,0) circle (0.4cm);
            \begin{feynman}
				\vertex (o) at (0,0) {\scalebox{.65}{$\widehat{R}^{(2)}$}};
			\end{feynman}
	\end{tikzpicture} }\,+\, \mathord{
\begin{tikzpicture}[scale=0.8,baseline=-0.65ex]
\begin{feynman}
	\vertex (O) at (-1,1); 
	\vertex (B) at (1.2,0);
	\vertex (A) at (2.4,0);
	\vertex (OO) at (0,0);
	\vertex (OOO) at (-1,0);
    \vertex (V) at (-0.9,0.3)  {\footnotesize{$i$}};
	\diagram*{
		(OOO)--[thick,dashed] (OO),
		(OO)--[plain, half left] (B),
		(OO)--[plain, half right] (B),
		(B)--[plain, half left] (A),
		(B)--[plain, half right ] (A),
	};
\end{feynman}
\filldraw[fill=red!20, draw=black] (0,0) circle (0.4cm);
\filldraw[fill=blue!20, draw=black] (1.2,0) circle (0.4cm);
\begin{feynman}
				\vertex (o) at (0,0) {\scalebox{.65}{$\widehat{R}^{(2)}$}};
                \vertex (oo) at (1.2,0) {\scalebox{.65}{$\widehat{Q}^{(4)}$}};
			\end{feynman}
\end{tikzpicture} }
    \,+\, \mathord{
		\begin{tikzpicture}[scale=0.8,baseline=-0.65ex]
			\begin{feynman}
				\vertex (O) at (0,0);
                \vertex (OO) at (0,1);
                 \vertex (OOO) at (0,-1);
				\vertex (A) at (0,0); 
				\vertex (AA) at (-1,1);
				\vertex (AAA) at (1,1);
				\vertex (B) at (-1.,0) ;
				\vertex (C) at (1.,0) ;
				\vertex (V) at (-0.9,0.3)  {\footnotesize{$i$}};
				\diagram*{
					(O) -- [plain, half left] (OO),
                    (O) -- [plain, half right ] (OO),
						(B) -- [thick,dashed] (O),
                        (O) -- [plain, half left] (OOO),
                    (O) -- [plain, half right ] (OOO),
				};
			\end{feynman}
			\filldraw[fill=red!20, draw=black] (0,0) circle (0.4cm);
            \begin{feynman}
				\vertex (o) at (0,0) {\scalebox{.65}{$\widehat{R}^{(4)}$}};
			\end{feynman}
	\end{tikzpicture} }+\ldots \ .
\end{equation}
Substituting this expression into (\ref{WtoG}) yields the large-$N$ expansion of $\langle \mathcal{W}\rangle_\Dhat$. Owing to the scaling properties of the vertices $\widehat{Q}^{(2m)}$ and $\widehat{R}^{(2m)}$, only a finite number of diagrams contribute at any fixed order in the large-$N$ expansion.

If all external sources $J_i$ are switched off, the above construction provides the Wilson loop expectation value in the \textbf{E}-theory. In practice, this corresponds to replacing $\widehat{Q}^{(2m)}$ with $Q^{(0|2m)}$ and $\widehat{R}^{(2m)}$ with $Q^{(1|2m)}$. In this way, we immediately recover the results of \cite{Beccaria:2021vuc,Beccaria:2022ypy,Beccaria:2023kbl} at the first few orders in the large-$N$ expansion. 

\section{The free energy at strong coupling}
\label{secn:free energy}

In this section we analyze the free energy of the $\widehat{\mathbf{D}}$-theory in the strong-coupling limit $\lambda\to\infty$. Building on the general framework developed in Section \ref{subsec:freeenergy}, we examine separately the various contributions appearing in (\ref{eq:FD1}) and determine their leading strong-coupling behavior to all orders in the large-$N$ expansion. Subleading effects in the conformal \textbf{D}-theory will be discussed in a subsequent subsection.

\subsection{Leading terms}
\label{subscn:freelead}

The $\mathcal{N}=4$ free energy is known exactly as a function of $\lambda$ and is reported in (\ref{FN=4}).
We therefore focus on the other terms in (\ref{eq:FD1}). 

\paragraph{The function $\widehat{Q}^{(0)}$:} The first contribution we have to consider is the large-$N$ expansion of $-\widehat{Q}^{(0)}$ given in (\ref{Qhat0exp1}). Its dependence on the 't Hooft coupling $\lambda$ is encoded in the coefficients $q_\ell$, defined in (\ref{eq:qellapp}) and originally expressed as perturbative series in $\lambda$. These series, however, can be resummed in terms of Mellin-Barnes integrals which can then be used to derive their asymptotic expansion at strong coupling. 

As an illustrative example, let us consider $q_1$. Its weak-coupling expansion (\ref{q1explicit}) admits the following Mellin-Barnes representation
\begin{align}
    q_1=\int_{\delta-\ii\infty}^{\delta+\ii\infty}\frac{ds}{2\pi\ii}\,\Big(\frac{\lambda}{16\pi^2}\Big)^{s+1}\frac{2\pi}{\sin\pi s}\,\frac{\Gamma(2s+2)\,\zeta(2s+1)}{\Gamma(s+2)\,\Gamma(s+3)}\,\big(2^{2s+2}-4+b_0\big)
    \label{q1s}
\end{align}
where $\delta\in (0,1)$. This representation is valid for all values of $\lambda$ and reduces to the original perturbative expression (\ref{q1explicit}) if  one closes the integration contour clockwise. If instead one closes the contour counter-clockwise, one obtains the following strong-coupling asymptotic expansion
\begin{align}
   q_1&\underset{\lambda \rightarrow \infty}{\sim} \, b_0\bigg[\frac{\lambda}{32\pi^2}\log \Big(\frac{\lambda }{16 \pi ^2}\rme^{2 \gamma }\Big)-\frac{\lambda}{64 \pi ^2}
    +\frac{1}{12}\log \Big(\frac{\lambda }{16 \pi^2}\Big)+\frac{\pi ^2}{15\lambda}\bigg]\notag\\[2mm]
    &\quad\quad~+\frac{\log 2}{4 \pi ^2}\lambda
    -\frac{1}{4}
   \log \Big(\frac{\lambda }{16 \pi ^2}\Big)-\frac{\pi ^2}{4 \lambda }+\text{constant}~.
   \label{q1strongfull}
\end{align}
Remarkably, this expansion truncates after finitely many terms, up to exponentially suppressed corrections.
The terms in the second line reproduce the results of \cite{Beccaria:2021ism}, while those in the first line constitute a new contribution.

Retaining only the leading terms as $\lambda\to\infty$, we have
\begin{align}
    q_1\underset{\lambda \rightarrow \infty}{\sim} \,\frac{\mathcal{Y}}{2}+\ldots
\label{q1strong}
\end{align}
where
\begin{subequations}
\begin{numcases}
{\mathcal{Y}=}
\frac{\log 2}{2\pi^2} \lambda
& \text{if } $b_0 = 0$~,\label{Yconf}
\\[3mm]
\frac{b_0}{16\pi^2}\lambda\log\lambda
& \text{if } $b_0 \neq 0$~. \label{Ynonconf}
\end{numcases}\label{Ystrong}%
\end{subequations}
A key observation is that this structure persists for all higher-order coefficients. Indeed, as shown in Appendix~\ref{App:leading}, the functions
$\mathcal{Y}_a$ defined in (\ref{Yell}) satisfy
\begin{align}
   \mathcal{Y}_a~ \underset{\lambda \rightarrow \infty}{\sim} \, -\mathcal{Y}+\ldots \qquad\mbox{for all}~a
   \label{Yastrong}
\end{align}
both in the conformal ($b_0=0$) and non-conformal ($b_0\neq0$) cases. As a consequence, the coefficients $q_\ell$ admit the following simple asymptotic behavior
\begin{align}
    q_\ell&\underset{\lambda \rightarrow \infty}{\sim} \,-(-1)^{\ell}\,\frac{\mathcal{Y}^\ell}{2\ell}+\ldots~.
    \label{qellstrong}
\end{align}
The derivation of this result is presented in Appendix~\ref{App:leading}.

Substituting (\ref{qellstrong}) into the expansion (\ref{Qhat0exp1}) yields
\begin{align}
    -\widehat{Q}^{(0)}&\underset{\lambda \rightarrow \infty}{\sim} ~N\Big(\frac{\mathcal{Y}}{2}+\ldots\Big)+\Big(\!\!-\frac{\mathcal{Y}^2}{4}+\ldots\Big)+
    \frac{1}{N}\Big(\frac{\mathcal{Y}^3}{6}+\ldots\Big)
    +\ldots~.
    \label{Qhat0final0}
\end{align}
In the regime 
 \begin{align}
\begin{cases}
    \!\!&1\ll\lambda\ll N \quad\mbox{if}~~b_0=0~,\\[2mm]
    \!\!&1\ll\lambda\ll\lambda\log\lambda\ll N\quad\mbox{if}~~b_0\not=0~,
\end{cases}
\label{regimes}
\end{align} 
or, equivalently $\frac{\mathcal{Y}}{N}\ll 1$, the series (\ref{Qhat0final0}) can be resummed to give
\begin{align}
    -\widehat{Q}^{(0)}\underset{\lambda \rightarrow \infty}{\sim} \frac{N^2}{2}\,\log\left(1+\frac{\mathcal{Y}}{N}\right)+\ldots~.
    \label{Qhat0final}
\end{align}
Combining this result with (\ref{FN=4}), and neglecting subleading terms discussed later\,\footnote{In particular, the $\frac{1}{2}\log\lambda$ term in (\ref{FN=4}) is subleading compared to the $O(N^0)$ contribution in (\ref{Qhat0final0}).}, we find
\begin{align}
    {F}_\NN-\widehat{Q}^{(0)}\underset{\lambda \rightarrow \infty}{\sim}
    -N^2\,\frac{\log \widetilde{\lambda}}{2}+\ldots
    \label{F-Qhat0}
\end{align}
where we introduced the rescaled coupling $\widetilde{\lambda}$ defined by
\begin{subequations}
\begin{numcases}
{\widetilde{\lambda}= \dfrac{\lambda}{1+\frac{\mathcal{Y}}{N}}=}
~\frac{\lambda}{1+\frac{\log 2}{2\pi^2 N}\lambda}
& \text{if } $b_0 = 0$~,\label{tildelambdaconf}
\\[3mm]
~\frac{\lambda}{1+\frac{b_0}{16\pi^2 N}\,\lambda\log\lambda}
& \text{if } $b_0 \neq 0$~. \label{tildelambdanonconf}
\end{numcases}\label{lambdatilde}%
\end{subequations}
Thus, at leading order, the sole effect of the contribution $-\widehat{Q}^{(0)}$ is simply to replace $\lambda$ by $\widetilde{\lambda}$ in the free energy of $\mathcal{N}=4$ SYM.

\paragraph{Trace-log and vacuum diagrams:}
The replacement mechanism exhibited in (\ref{F-Qhat0}) extends to the remaining contributions in (\ref{eq:FD1}), namely the trace-log term and the vacuum diagrams involving the interaction vertices $\widehat{Q}^{(2m)}$ with $m\ge1$. As shown in Appendix~\ref{sec:properties}, the strong-coupling behavior of these vertices is governed by the asymptotic relation
\begin{equation}
    \label{hatQ2m}
      \widehat{Q}^{(2m)}_{j_1,\ldots,j_{2m}} \underset{\lambda \rightarrow \infty}{\sim}~
         \frac{1}{\left(1 + \frac{\cY}{N}\right)^{\sum_k j_k + m }}\, Q^{(0|2m)}_{j_1,\ldots, j_{2m}}~,
\end{equation} 
where $Q^{(0|2m)}_{j_1,\ldots, j_{2m}}$ are the connected Gaussian correlators appearing in the effective action (\ref{ZeffJE}) of the $\mathbf{E}$-theory. We now demonstrate that this relation implies that, at any fixed order in the $1/N$-expansion, the dominant strong-coupling contributions can be expressed in terms of $\mathbf{E}$-theory diagrams evaluated at the rescaled coupling $\widetilde{\lambda}$.

To see this, let us first consider the trace-log term \eqref{trlogexpanded}
\begin{align}
    \label{trlogexpanded1}
        \frac{1}{2}\tr \log\big(1- C\,\widehat{Q}^{(2)}\big)&= -\sum_{k=1}^\infty \frac{1}{2k}\bigg(\prod_{p=0}^{k-1} C_{j_{2p}\,j_{2p+1}}\,\widehat{Q}^{(2)}_{j_{2p+1}\,j_{2p+2}}\bigg)~,
\end{align} 
and recall from (\ref{eq:C^-}) that
\begin{align}
    C_{j_{2p}\,j_{2p+1}}=\lambda^{j_{2p}+j_{2p+1}+1}\,\widehat{C}_{j_{2p}\,j_{2p+1}}
    \label{Chatjj}
\end{align}
with $\widehat{C}$ independent of $\lambda$. Using this scaling behavior together with (\ref{hatQ2m}) for $m=1$, one finds that at strong coupling all explicit factors of $\lambda$ recombine into powers of $\widetilde{\lambda}$. Indeed
\begin{align}
\frac{1}{2}\tr \log\big(1-C\,\widehat{Q}^{(2)}\big)
&\underset{\lambda \rightarrow \infty}{\sim}
-\sum_{k=1}^\infty\frac{1}{2k}
\Bigg[\prod_{p=0}^{k-1}\,
\widetilde{\lambda}^{\,j_{2p}+j_{2p+1}+1}\,
\widehat{C}_{j_{2p}\,j_{2p+1}}\,Q^{(0|2)}_{j_{2p+1}\,j_{2p+2}}\Bigg]+\ldots\notag\\[2mm]
&\underset{\lambda \rightarrow \infty}{\sim}~\frac{1}{2}\tr\log\big(1-C\,Q^{(0|2)}\big)\Big|_{\lambda\to\widetilde{\lambda}}
+\ldots~.
\label{trlogexpanded2}
\end{align}
This is precisely the trace-log contribution to the free energy of the \textbf{E}-theory evaluated at coupling $\widetilde{\lambda}$.
Exploiting the results of~\cite{Beccaria:2023kbl}, one therefore obtains
\begin{align}
\frac{1}{2}\tr\log\big(1-C\,\widehat{Q}^{(2)}\big)
\underset{\lambda \rightarrow \infty}{\sim}
\frac{\sqrt{\widetilde{\lambda}}}{8}+\ldots.~
\label{trlogexpanded2a}
\end{align}
An independent check follows from inserting the leading strong-coupling behavior (\ref{Yastrong}) of the functions $\mathcal{Y}_a$, together with the large-$\lambda$ asymptotics of the matrix elements $w_{n,m}$ derived in \cite{Beccaria:2023kbl}, into the large-$N$ expansion (\ref{trloglargeN}). This yields
\begin{align}
    \frac{1}{2}\tr \log\big(1- C\,\widehat{Q}^{(2)}\big)~&\underset{\lambda \rightarrow \infty}{\sim}\Big(\frac{\sqrt{\lambda}}{8}+\ldots\Big)+
    \frac{1}{N}\Big(\!\!-\frac{\sqrt{\lambda}\,\mathcal{Y}}{16}+\ldots\Big)+\frac{1}{N^2}\Big(\frac{3\sqrt{\lambda}\,\mathcal{Y}^2}{64}+\ldots\Big)+\ldots\notag\\[2mm]~&\underset{\lambda \rightarrow \infty}{\sim}    \frac{\sqrt{\widetilde{\lambda}}}{8}+\ldots \ ,
    \label{trlogexpl}
\end{align}
in agreement with (\ref{trlogexpanded2a}). Notice that these contributions are subleading compared to those generated by $-\widehat{Q}^{(0)}$ at the corresponding orders in the $1/N$-expansion.

The analysis of the remaining vacuum diagrams in~\eqref{eq:FD1} proceeds analogously. As a representative example, let us consider the diagram
\begin{equation}
    \label{1std}
    \begin{tikzpicture}[scale=0.8,baseline=-0.65ex]
        \begin{feynman}
        	\vertex (O) at (0,1); 
        	\vertex (OO) at (0,0);
        	\vertex (OOO) at (0,-1);			
        	\diagram*{
        		(OO)--[plain, half left,looseness=1.1] (OOO),
        		(OO)--[plain, half right,looseness=1.1] (OOO),
        			(OO)--[plain, half left,looseness=1.1] (O),
        		(OO)--[plain, half right,looseness=1.1] (O),
        	};
        \end{feynman}
        \filldraw[fill=blue!20, draw=black] (0,0) circle (0.4cm);
        \begin{feynman}
				\vertex (o) at (0,0) {\scalebox{.65}{$\widehat{Q}^{(4)}$}};
			\end{feynman}
        \end{tikzpicture}
        = -\frac{1}{8}\,\widehat{Q}^{(4)}_{j_1j_2j_3j_4}\, X^{j_1j_2}\, X^{j_3j_4}
\end{equation}
where $X$ is the propagator defined in (\ref{defXprop}). This propagator can be manipulated as follows 
\begin{align}
        X^{j_1j_2} & = \Big[C + C \,\widehat{Q}^{(2)} C +  C \,\widehat{Q}^{(2)} C \, \widehat{Q}^{(2)} C + \ldots \Big]^{j_1j_2}\notag\\[2mm]
        &\!\underset{\lambda \rightarrow \infty}{\sim}
        \big(1+\tfrac{\mathcal{Y}}{N}\big)^{j_1+j_2+1}\,\Bigg\{\!\Big[C + C\, Q^{(0|2)} C +  C\, Q^{(0|2)} C \, Q^{(0|2)} C + \ldots \Big]^{j_1j_2}\,\bigg|_{\lambda \to \widetilde{\lambda}}\,\Bigg\}\!+\ldots\notag\\[2mm]
        &\!\underset{\lambda \rightarrow \infty}{\sim}\big(1+\tfrac{\mathcal{Y}}{N}\big)^{j_1+j_2+1}\,X_\E^{j_1j_2}(\widetilde{\lambda})+\ldots    \label{Xsc} ~,
\end{align} 
where in the second line we applied (\ref{hatQ2m}) with $m=1$ and in the last step we identified the propagator of the \textbf{E}-theory with coupling $\widetilde{\lambda}$. 
Substituting (\ref{Xsc}) into (\ref{1std}) and using (\ref{hatQ2m}) with $m=2$, 
 we readily obtain 
 \begin{equation}
    \label{1stdstrong}
    \begin{tikzpicture}[scale=0.8,baseline=-0.65ex]
        \begin{feynman}
        	\vertex (O) at (0,1); 
        	\vertex (OO) at (0,0);
        	\vertex (OOO) at (0,-1);			
        	\diagram*{
        		(OO)--[plain, half left,looseness=1.1] (OOO),
        		(OO)--[plain, half right,looseness=1.1] (OOO),
        			(OO)--[plain, half left,looseness=1.1] (O),
        		(OO)--[plain, half right,looseness=1.1] (O),
        	};
        \end{feynman}
        \filldraw[fill=blue!20, draw=black] (0,0) circle (0.4cm);
        \begin{feynman}
				\vertex (o) at (0,0) {\scalebox{.65}{$\widehat{Q}^{(4)}$}};
			\end{feynman}
        \end{tikzpicture}
        ~\underset{\lambda \rightarrow \infty}{\sim}~ -\frac{1}{8}\,{Q}^{(0|4)}_{j_1j_2j_3j_4}\, X_\E^{j_1j_2}(\widetilde{\lambda})\, X_\E^{j_3j_4}(\widetilde{\lambda})+\ldots~.
\end{equation}
The right-hand side is precisely the same diagram as the left-hand side, but evaluated in the \textbf{E}-theory at coupling $\widetilde{\lambda}$. Using the explicit results of \cite{Beccaria:2023kbl}, one finds 
\begin{align}
\begin{tikzpicture}[scale=0.8,baseline=-0.65ex]
        \begin{feynman}
        	\vertex (O) at (0,1); 
        	\vertex (OO) at (0,0);
        	\vertex (OOO) at (0,-1);			
        	\diagram*{
        		(OO)--[plain, half left,looseness=1.1] (OOO),
        		(OO)--[plain, half right,looseness=1.1] (OOO),
        			(OO)--[plain, half left,looseness=1.1] (O),
        		(OO)--[plain, half right,looseness=1.1] (O),
        	};
        \end{feynman}
        \filldraw[fill=blue!20, draw=black] (0,0) circle (0.4cm);
        \begin{feynman}
				\vertex (o) at (0,0) {\scalebox{.65}{$\widehat{Q}^{(4)}$}};
			\end{feynman}
        \end{tikzpicture}
        ~&\underset{\lambda \rightarrow \infty}{\sim}~
        -\frac{1}{N^2}\Big(\frac{1}{32}w_{0,0}^2+\frac{1}{32}w_{0,0}\,w_{0,1}\Big)\bigg|_{\lambda \to \widetilde{\lambda}}+\ldots
    =-\frac{1}{N^2}\Big(\frac{\widetilde{\lambda}^{\,3/2}}{2048}+\ldots\Big)+\ldots~.
    \label{vacuumQhat4}
\end{align}
These contributions are subleading with respect to those arising from $-\widehat{Q}^{(0)}$ and the trace-log term at the same order in $1/N$.

It is straightforward to verify that the same mechanism applies to all remaining vacuum diagrams in (\ref{eq:FD1}), and more generally to all diagrams contributing to the free energy. In particular, all factors of $(1+\frac{\mathcal{Y}}{N})$ that do not recombine with powers of $\lambda$ to reconstruct $\widetilde{\lambda}$ cancel identically. Consequently, at each order in the $1/N$ expansion, the leading strong-coupling contributions to the free energy of the $\widehat{\mathbf{D}}$-theory can be resummed into expressions depending only on the effective coupling $\widetilde{\lambda}$.
This holds both in the conformal case ($b_0=0$) and in the non-conformal case ($b_0\neq0$), the only difference being the explicit form of $\mathcal{Y}$ in (\ref{Ystrong}). Retaining only the dominant terms, we obtain
\begin{align}
    F_\Dhat \underset{\lambda \rightarrow \infty}{\sim}~-N^2\,\frac{\log\widetilde{\lambda}}{2}+\ldots
    \label{FDhatleading}
\end{align}
with $\widetilde{\lambda}$ given in (\ref{lambdatilde}). For non-conformal theories with $b_0\not=0$ this represents a genuinely new result.

In the following subsection we extend this analysis to the subleading terms in the conformal \textbf{D}-theory.

\subsection{Subleading terms}
\label{subsecn:subleading}

When $b_0=0$, the strong-coupling expansion of the free energy can be systematically extended to subleading orders using the same methods discussed above\,\footnote{For the non-conformal theories, the subleading contributions have a much more intricate structure due to interferences between terms proportional to $b_0$ with those already present in the conformal case. Although it is possible to compute some specific contributions, writing general formulas is highly non-trivial and thus we have not attempted in this paper to perform this analysis in a systematic way.}. The detailed derivation is presented in Appendix~\ref{App:subleading}; here we summarize the main results.

\paragraph{The function $\widehat{Q}^{(0)}$:} Let us consider the large-$N$ expansion of $-\widehat{Q}^{(0)}$ given in (\ref{Qhat0exp1}). In the strong-coupling limit $\lambda\to\infty$, the leading coefficient $q_1$ behaves as 
\begin{align}
    q_1\,&\underset{\lambda \rightarrow \infty}{\sim}~\frac{\mathcal{Y}}{2}-\frac{\log\lambda}{4}+O\big(\lambda^0\big)~,
    \label{q1string2}
\end{align}
with $\mathcal{Y}$ given in (\ref{Yconf}), as follows from the second line of (\ref{q1strongfull}). The asymptotic behavior of the higher coefficients is controlled by that of the functions 
$\mathcal{Y}_a$, which take the universal form (see (\ref{eq:Y1strong}))
\begin{align}
    \mathcal{Y}_a~\underset{\lambda \rightarrow \infty}{\sim}~-\mathcal{Y}+\frac{1}{4}\,\delta_{a,1}
    \label{Yastrongexact}
\end{align}
up to exponentially suppressed contributions. Using this result, one finds (see (\ref{q0nfinal}))
\begin{subequations}
     \begin{align}
    \label{q1q2sc}
        q_2\,&\underset{\lambda \rightarrow \infty}{\sim}~-\frac{\mathcal{Y}^2}{4}+\frac{\mathcal{Y}}{4}-\frac{\log\lambda}{32}+O\big(\lambda^0\big)~,\\[2mm]
        q_\ell\,&\underset{\lambda \rightarrow \infty}{\sim}~-(-1)^{\ell}\,\frac{\mathcal{Y}^\ell}{2\ell}+(-1)^{\ell}\frac{\mathcal{Y}^{\ell-1}}{4(\ell-1)}
        +(-1)^{\ell}\frac{15\,\mathcal{Y}^{\ell-2}}{32(\ell-2)}+O\big(\lambda^0\big)\quad\mbox{for}~\ell\geq3~.
      \label{eq:qellsub}
    \end{align}
\label{qellsub}%
\end{subequations}
Importantly, the $\log\lambda$-terms are present only in $q_1$ and $q_2$. This implies that the logarithmic corrections appear only up to $O(N^0)$ in the large-$N$ expansion of the free energy.

Substituting (\ref{q1string2}) and (\ref{qellsub}) into (\ref{Qhat0exp1}), and combining the results with the free energy (\ref{FN=4}) of $\mathcal{N}=4$ SYM, we find that all the subleading contributions can be resummed  in terms of the rescaled coupling $\widetilde{\lambda}$ defined in (\ref{tildelambdaconf}). Indeed, a straightforward calculation shows that 
\begin{align}
    \label{Fn4q1}
        {F}_\NN-\widehat{Q}^{(0)}\underset{\lambda \rightarrow \infty}{\sim}
        -N^2\,\frac{\log \widetilde{\lambda}}{2}-N\,\frac{\log \widetilde{\lambda}}{4}+\frac{15}{32}\log \widetilde{\lambda}+O\big(\widetilde{\lambda}^0\big)~,
    \end{align}
which extends (\ref{F-Qhat0}) to subleading orders in the conformal \textbf{D}-theory.

\paragraph{Trace-log and vacuum diagrams:} We now turn our attention to the trace-log term, whose large-$N$ expansion is given in (\ref{trloglargeN}). Using (\ref{Yastrongexact}), together with the strong-coupling behavior of $\tr \log(1- C\,Q^{(0|2)})$ and of the matrix elements $w_{n,m}$ derived in \cite{Beccaria:2023kbl}, we obtain
\begin{align}
    \frac{1}{2}\tr \log\big(1- C\,\widehat{Q}^{(2)}\big)&\underset{\lambda \rightarrow \infty}{\sim}\Big(\frac{\sqrt{\lambda}}{8}-\frac{3\log\lambda}{8}+O(\lambda^0)\Big)\!+\!
    \frac{1}{N}\Big(\!\!-\frac{\sqrt{\lambda}\,\mathcal{Y}}{16}+\frac{3\mathcal{Y}}{8}+\frac{\sqrt{\lambda}}{64}+O(\lambda^0)\Big)\notag\\[2mm]
    &\hspace{-2.5cm}+\frac{1}{N^2}\Big(\frac{3\sqrt{\lambda}\,\mathcal{Y}^2}{64}-\frac{3\mathcal{Y}^2}{16}-\frac{\sqrt{\lambda}\,\mathcal{Y}}{128}-\frac{9\lambda}{2048}-\frac{(3-2\log 2)\sqrt{\lambda}}{2048}+O(\lambda^0)\Big)+\ldots~.\label{trlogsubleading}
\end{align}
All dependence on $\mathcal{Y}$ can again be reabsorbed into the rescaled coupling $\widetilde{\lambda}$, allowing us to rewrite this expression as 
\begin{align}
    \frac{1}{2}\tr \log\big(1- C\,\widehat{Q}^{(2)}\big)&\underset{\lambda \rightarrow \infty}{\sim}
\bigg(\!\frac{\sqrt{\widetilde{\lambda}}}{8}-\frac{3\log\widetilde{\lambda}}{8}\!\bigg)+
    \frac{1}{N}\frac{\sqrt{\widetilde{\lambda}}}{64}\notag\\[2mm]
&\qquad~-\frac{1}{N^2}\bigg(\frac{9\widetilde{\lambda}}{2048}+\frac{(3-2\log 2)\sqrt{\widetilde{\lambda}}}{2048}\bigg)+O(\widetilde{\lambda}^0)~.
\end{align}

The remaining contribution at this order arises from the vacuum diagram involving the vertex $\widehat{Q}^{(4)}$ shown in (\ref{Q4diagram}). Using the results of \cite{Beccaria:2023kbl}, we obtain
\begin{align}
    \begin{tikzpicture}[scale=0.8,baseline=-0.65ex]
        \begin{feynman}
        	\vertex (O) at (0,1); 
        	\vertex (OO) at (0,0);
        	\vertex (OOO) at (0,-1);			
        	\diagram*{
        		(OO)--[plain, half left,looseness=1.1] (OOO),
        		(OO)--[plain, half right,looseness=1.1] (OOO),
        			(OO)--[plain, half left,looseness=1.1] (O),
        		(OO)--[plain, half right,looseness=1.1] (O),
        	};
        \end{feynman}
        \filldraw[fill=blue!20, draw=black] (0,0) circle (0.4cm);
        \begin{feynman}
				\vertex (o) at (0,0) {\scalebox{.65}{$\widehat{Q}^{(4)}$}};
			\end{feynman}
        \end{tikzpicture}
        \underset{\lambda \rightarrow \infty}{\sim}-\frac{1}{N^2}\bigg(\frac{\widetilde{\lambda}^{3/2}}{2048}-\frac{3\widetilde{\lambda}}{1024}
        -\frac{3(2-3\log 2)\sqrt{\widetilde{\lambda}}}{1024}\bigg)+O(\widetilde{\lambda}^0)~.
\end{align}

Collecting all terms, we find that the strong-coupling expansion of the free energy of the \textbf{D}-theory up to order $1/N^2$ is
\begin{align}
    F_\D&\underset{\lambda \rightarrow \infty}{\sim}-N^2\,\frac{\log \widetilde{\lambda}}{2}-N\,\frac{\log \widetilde{\lambda}}{4}+\bigg(\frac{\sqrt{\widetilde{\lambda}}}{8}+\frac{3\log \widetilde{\lambda}}{32}\bigg)+
    \frac{1}{N}\frac{\sqrt{\widetilde{\lambda}}}{64}\notag\\[2mm]
&\qquad~-\frac{1}{N^2}\bigg(\frac{\widetilde{\lambda}^{3/2}}{2048}+\frac{3\widetilde{\lambda}}{2048}-\frac{(9-16\log 2)\sqrt{\widetilde{\lambda}}}{2048}\bigg)+O(\widetilde{\lambda}^0)~.
\label{FDstrong}
\end{align}
We observe that leading and subleading terms in this expression are
\begin{align}
    -2\Big(\frac{N^2}{4}+\frac{N}{8}\Big)\log\widetilde{\lambda}
\end{align}
which encode the $N$-dependent part of the $a$-anomaly coefficient of the \textbf{D}-theory\,\footnote{The $a$-anomaly coefficient is determined by the matter content and for the \textbf{D}-theory is $a=\frac{N^2}{4}+\frac{N}{8}-\frac{5}{24}$. See also the comments in footnote \ref{footnote_intro}.}. This is a nice feature of our results.

Finally, it is instructive to compare $F_\D$ with the free energy of the \textbf{E}-theory at strong coupling derived in \cite{Beccaria:2023kbl}, which reads
\begin{align}
    F_\E&\underset{\lambda \rightarrow \infty}{\sim}-N^2\,\frac{\log{\lambda}}{2}+\bigg(\frac{\sqrt{{\lambda}}}{8}+\frac{\log {\lambda}}{8}\bigg)+\notag\\[2mm]
&\qquad~-\frac{1}{N^2}\bigg(\frac{{\lambda}^{3/2}}{2048}+\frac{3{\lambda}}{2048}-\frac{(11-16\log 2)\sqrt{{\lambda}}}{2048}\bigg)+O({\lambda}^0)~.
\label{FEstrong}
\end{align}
A direct comparison shows that
\begin{align}
\label{FDFE}
    F_\D&\underset{\lambda \rightarrow \infty}{\sim} ~F_\E\Big|_{\lambda\to\widetilde{\lambda}}-N\,\frac{\log \widetilde{\lambda}}{4}-\frac{\log \widetilde{\lambda}}{32}+\frac{1}{N}\frac{\sqrt{\widetilde{\lambda}}}{64}-\frac{1}{N^2}\frac{\sqrt{\widetilde{\lambda}}}{1024}+O(\widetilde{\lambda}^0)~,
\end{align}
which makes explicit the precise relation between the free energies of the \textbf{D}- and \textbf{E}-theories at strong coupling.

\section{The circular Wilson loop at strong coupling}
\label{secn:Wilson loop}
We now turn our attention to the strong-coupling behavior of the vacuum expectation value of the $\frac{1}{2}$-BPS Wilson loop $\mathcal{W}$ in the large-$N$ limit. Combining (\ref{WtoG}) with the diagrammatic representation (\ref{O2ndiag}), we find that $\big\langle \mathcal{W} \big\rangle_\Dhat$ admits the following expansion
\begin{align}
    \label{WtoG1}
        \big\langle \mathcal{W} \big\rangle_\Dhat
        = 1 + \frac{1}{N}\sum_{i=1}^\infty\frac{1}{(2i)!}\,\Big(\frac{\lambda}{2}\Big)^i
        \Bigg[
        & \mathord{
		\begin{tikzpicture}[scale=0.8,baseline=-0.65ex]
			\begin{feynman}
				\vertex (O) at (0,0);
				\vertex (AA) at (-1,1);
				\vertex (AAA) at (1,1);
				\vertex (B) at (-1.,0) ;
                \vertex (D) at (-0.9,0.3)  {\footnotesize{$i$}};
				\diagram*{
						(B) -- [thick,dashed] (O),
				};
			\end{feynman}
			\filldraw[fill=red!20, draw=black] (0,0) circle (0.4cm);
            \begin{feynman}
				\vertex (o) at (0,0) {\scalebox{.65}{$\widehat{R}^{(0)}$}};
			\end{feynman}
	\end{tikzpicture} }       
        \,+\, 	\mathord{
		\begin{tikzpicture}[scale=0.8,baseline=-0.65ex]
			\begin{feynman}
				\vertex (O) at (0,0);
                \vertex (OO) at (1.2,0);
				\vertex (A) at (0,0); 
				\vertex (AA) at (-1,1);
				\vertex (AAA) at (1,1);
				\vertex (B) at (-1.,0) ;
                \vertex (D) at (-0.9,0.3)  {\footnotesize{$i$}};
				\vertex (C) at (1.,0) ;
				\diagram*{
					(O) -- [plain, half left] (OO),
                    (O) -- [plain, half right ] (OO),
						(B) -- [thick,dashed] (O),
				};
			\end{feynman}
			\filldraw[fill=red!20, draw=black] (0,0) circle (0.4cm);
            \begin{feynman}
				\vertex (o) at (0,0) {\scalebox{.65}{$\widehat{R}^{(2)}$}};
			\end{feynman}
	\end{tikzpicture} }\,
    \nonumber\\
    & +\, \mathord{
\begin{tikzpicture}[scale=0.8,baseline=-0.65ex]
\begin{feynman}
	\vertex (O) at (-1,1); 
	\vertex (B) at (1.2,0);
	\vertex (A) at (2.4,0);
	\vertex (OO) at (0,0);
	\vertex (OOO) at (-1,0);
    \vertex (V) at (-0.9,0.3)  {\footnotesize{$i$}};
	\diagram*{
		(OOO)--[thick,dashed] (OO),
		(OO)--[plain, half left] (B),
		(OO)--[plain, half right] (B),
		(B)--[plain, half left] (A),
		(B)--[plain, half right ] (A),
	};
\end{feynman}
\filldraw[fill=red!20, draw=black] (0,0) circle (0.4cm);
\filldraw[fill=blue!20, draw=black] (1.2,0) circle (0.4cm);
\begin{feynman}
				\vertex (o) at (0,0) {\scalebox{.65}{$\widehat{R}^{(2)}$}};
                \vertex (oo) at (1.2,0) {\scalebox{.65}{$\widehat{Q}^{(4)}$}};
			\end{feynman}
\end{tikzpicture} }
    \,+\, \mathord{
		\begin{tikzpicture}[scale=0.8,baseline=-0.65ex]
			\begin{feynman}
				\vertex (O) at (0,0);
                \vertex (OO) at (0,1);
                 \vertex (OOO) at (0,-1);
				\vertex (A) at (0,0); 
				\vertex (AA) at (-1,1);
				\vertex (AAA) at (1,1);
				\vertex (B) at (-1.,0) ;
				\vertex (C) at (1.,0) ;
				\vertex (V) at (-0.9,0.3)  {\footnotesize{$i$}};
				\diagram*{
					(O) -- [plain, half left] (OO),
                    (O) -- [plain, half right ] (OO),
						(B) -- [thick,dashed] (O),
                        (O) -- [plain, half left] (OOO),
                    (O) -- [plain, half right ] (OOO),
				};
			\end{feynman}
			\filldraw[fill=red!20, draw=black] (0,0) circle (0.4cm);
            \begin{feynman}
				\vertex (o) at (0,0) {\scalebox{.65}{$\widehat{R}^{(4)}$}};
			\end{feynman}
	\end{tikzpicture} }+\ldots 
        \Bigg]~,
\end{align} 
where the functions $\widehat{R}^{(2m)}$ are defined in (\ref{defRtilde}) and depend on the interaction vertices $\widehat{Q}^{(2m)}$ (\ref{Qhat}), which in turn are expressed in terms of the Gaussian correlators $Q^{(n|2m)}$ given in (\ref{defQs}). Using the topological expansion (\ref{expansionmain}), each diagram in (\ref{WtoG1}) can be systematically expanded in powers of $1/N$, leading to the general structure  
\begin{equation}
\label{largeW}
    \big\langle \mathcal{W} \big\rangle_\Dhat = \sum_{n=0}^\infty \frac{1}{N^n}\mathcal{W}_{n} (\lambda)~.
\end{equation} 
The coefficients $\mathcal{W}_{n}(\lambda)$ receive contributions from different classes of diagrams. In particular, the diagram involving
$\widehat{R}^{(0)}$ contributes to all orders in $1/N$, whereas those containing $\widehat{R}^{(2)}$ start contributing at order $1/N^2$. Higher vertices behave similarly.  It is therefore convenient to study each diagram separately and decompose the Wilson loop expectation value as
\begin{align}
     \big\langle \mathcal{W} \big\rangle_\Dhat= \widehat{\mathcal{W}}^{(0)}+\widehat{\mathcal{W}}^{(2)}+\ldots
\end{align}
where
\begin{subequations}
\begin{align}
    \widehat{\mathcal{W}}^{(0)}&=
    1+\frac{1}{N}\sum_{i=1}^\infty\frac{1}{(2i)!}\,\Big(\frac{\lambda}{2}\Big)^i\bigg(
      \mathord{
	\begin{tikzpicture}[scale=0.8,baseline=-0.65ex]
		\begin{feynman}
				\vertex (O) at (0,0);
				\vertex (AA) at (-1,1);
				\vertex (AAA) at (1,1);
				\vertex (B) at (-1.,0) ;
                \vertex (D) at (-0.9,0.3)  {\footnotesize{$i$}};
				\diagram*{
						(B) -- [thick,dashed] (O),
				};
			\end{feynman}
			\filldraw[fill=red!20, draw=black] (0,0) circle (0.4cm);
            \begin{feynman}
				\vertex (o) at (0,0) {\scalebox{.65}{$\widehat{R}^{(0)}$}};
			\end{feynman}
	\end{tikzpicture} } \bigg)~,\label{wn4Jdef}\\[2mm]
  \widehat{\cW}^{(2)}
        & =\frac{1}{N}\sum_{i=1}^\infty\frac{1}{(2i)!} \Big(\frac{\lambda}{2}\Big)^i\bigg(  \mathord{
		\begin{tikzpicture}[scale=0.8,baseline=-0.65ex]
			\begin{feynman}
				\vertex (O) at (0,0);
                \vertex (OO) at (1.2,0);
				\vertex (A) at (0,0); 
				\vertex (AA) at (-1,1);
				\vertex (AAA) at (1,1);
				\vertex (B) at (-1.,0) ;
                \vertex (D) at (-0.9,0.3)  {\footnotesize{$i$}};
				\vertex (C) at (1.,0) ;
				\diagram*{
					(O) -- [plain, half left] (OO),
                    (O) -- [plain, half right ] (OO),
						(B) -- [thick,dashed] (O),
				};
			\end{feynman}
			\filldraw[fill=red!20, draw=black] (0,0) circle (0.4cm);
            \begin{feynman}
				\vertex (o) at (0,0) {\scalebox{.65}{$\widehat{R}^{(2)}$}};
			\end{feynman}
	\end{tikzpicture} }\bigg)~.\label{what1}
\end{align}%
\end{subequations}
Higher-order contributions can be treated in complete analogy. Although these expressions are initially given as perturbative series in $\lambda$, their strong-coupling behavior can be extracted using the same methods developed for the free energy.

\paragraph{The function $\widehat{\mathcal{W}}^{(0)}$:}
From (\ref{defRtilde}), we see that the vertex $\widehat{R}^{(0)}$ takes the form
\begin{align}
\label{R0is}
  \mathord{
		\begin{tikzpicture}[scale=0.8,baseline=-0.65ex]
			\begin{feynman}
				\vertex (O) at (0,0);
				\vertex (AA) at (-1,1);
				\vertex (AAA) at (1,1);
				\vertex (B) at (-1.,0) ;
                \vertex (D) at (-0.9,0.3)  {\footnotesize{$i$}};
				\diagram*{
						(B) -- [thick,dashed] (O),
				};
			\end{feynman}
			\filldraw[fill=red!20, draw=black] (0,0) circle (0.4cm);
            \begin{feynman}
				\vertex (o) at (0,0) {\scalebox{.65}{$\widehat{R}^{(0)}$}};
			\end{feynman}
	\end{tikzpicture} } \,\equiv\,
\widehat{R}^{(0)}_i=Q^{(1|0)}_{i}+ J_{k_1}\,Q^{(2|0)}_{i,k_1}  + \frac{1}{2}J_{k_1} J_{k_2} Q^{(3|0)}_{i,k_1k_2} +  \ldots~.
\end{align}
The leading term $Q^{(1|0)}_{i}$ is independent of $\lambda$ while the other terms depend on the 't Hooft coupling through the sources $J_k$ (\ref{eq:source K}). 
We now show that the contribution of $Q^{(1|0)}_{i}$ to (\ref{wn4Jdef}) reproduces the Wilson loop expectation value of $\mathcal{N}=4$ SYM at all orders in $1/N$. Indeed, using the large-$N$ expansion (see Appendix~\ref{sec:GC}), 
\begin{align}
    Q^{(1|0)}_i=N\,\beta_i\Big[\frac{1}{i(i+1)}+\frac{1}{N^2}\frac{i-7}{12}+O(1/N^4)\Big]~,
\end{align}
one finds
\begin{align}
1+\frac{1}{N}\sum_{i=1}^\infty\frac{1}{(2i)!} \Big(\frac{\lambda}{2}\Big)^i\, Q^{(1|0)}_{i} 
    &=\frac{2\,I_1\big(\sqrt{\lambda}\big)}{\sqrt{\lambda}}+\frac{1}{N^2}\frac{\lambda \,I_0\big(\sqrt{\lambda}\big)-14\sqrt{\lambda}\,I_1\big(\sqrt{\lambda}\big)}{48}+O(1/N^4)
    \label{N4a}
\end{align}
where $I_n$ are the modified Bessel functions of the first kind. 
These are precisely the planar and subplanar contributions to the Wilson loop expectation value in $\mathcal{N}=4$ SYM \cite{Erickson:2000af,Drukker:2000rr}. Extending the argument to all orders, we conclude that
\begin{align}
   1+\frac{1}{N}\sum_{i=1}^\infty\frac{1}{(2i)!} \Big(\frac{\lambda}{2}\Big)^i\, Q^{(1|0)}_{i} = \big\langle \mathcal{W}\big\rangle_\NN~.
     \label{N4}
\end{align}
Since no other diagrams contribute at order $N^0$, the planar term of $\big\langle\mathcal{W}\big\rangle_{\Dhat}$ coincides with that of $\mathcal{N}=4$ SYM:
\begin{align}
    \mathcal{W}_0(\lambda)=\lim_{N\to\infty} \big\langle \mathcal{W}\big\rangle_\NN=\frac{2\,I_1\big(\sqrt{\lambda}\big)}{\sqrt{\lambda}}~.
    \label{eq:W0}
\end{align}

While (\ref{N4a}) is exact in $\lambda$, its strong-coupling limit follows immediately from the large-argument expansion of the Bessel functions and turns out to be
\begin{equation}
        \big\langle \mathcal{W}\big\rangle_\NN~ \underset{\lambda \rightarrow \infty}{\sim} \sqrt{\frac{2}{\pi}}\, \frac{\ee^{\sqrt{\lambda}}}{\lambda^{3/4}} \Bigg[\bigg(1 -
        \frac{3}{8\sqrt{\lambda}} + \ldots\bigg) + \frac{1}{N^2} \bigg(\frac{\lambda^{3/2}}{96} + \ldots\bigg) + O(1/N^4)\Bigg]
         \label{N4ll}
\end{equation}
where we dropped exponentially suppressed terms.

We now consider the contributions to $\widehat{\mathcal{W}}^{(0)}$ arising from the $J$-dependent terms in (\ref{R0is}). 
Although these terms are more involved, their leading strong-coupling behavior drastically simplifies and, in the regime (\ref{regimes}), they can be resummed. Indeed, as shown in Appendix \ref{sec:properties}, one has
\begin{equation}
    \label{R0toQ1m}
         \widehat R^{(0)}_{i} \underset{\lambda \rightarrow \infty}{\sim} 
         \frac{1}{(1 + \tfrac{\cY}{N})^{i}}\, Q^{(1|0)}_{i}+\ldots
\end{equation}
with $\mathcal{Y}$ given in (\ref{Ystrong}). Substituting (\ref{R0toQ1m}) into (\ref{wn4Jdef}) and using (\ref{N4}), we obtain
\begin{align}
    \label{wn4J}
        \widehat{\mathcal{W}}^{(0)}
        \underset{\lambda \rightarrow \infty}{\sim}
        1 + \frac{1}{N}\sum_{i=1}^\infty\frac{1}{(2i)!} \bigg(\frac{\lambda}{2(1 + \tfrac{\cY}{N})}\bigg)^i\, Q^{(1|0)}_{i}+\ldots 
         \underset{\lambda \rightarrow \infty}{\sim}  ~  \big\langle \mathcal{W}\big\rangle_\NN\Big|_{\lambda\to\widetilde{\lambda}}~+\ldots ~.
\end{align}        
Thus, the leading strong-coupling contributions can be resummed to all orders in $1/N$ and expressed in terms of the $\mathcal{N}=4$  Wilson loop evaluated at the rescaled coupling $\widetilde\lambda$ defined in (\ref{lambdatilde}). This result closely mirrors the behavior of the free energy (\ref{F-Qhat0}).

Applying the replacement $\lambda\to\tilde\lambda$ to (\ref{N4ll}), we find
\begin{align}
    \label{N4llt}
      \widehat{\mathcal{W}}^{(0)}
       &\underset{\lambda \rightarrow \infty}{\sim}  \sqrt{\frac{2}{\pi}}\, \frac{\ee^{\sqrt{\lambda}}}{\lambda^{3/4}} \Bigg[\bigg(1 - \frac 38 
        \frac{1}{\sqrt{\lambda}} + \ldots\bigg) 
        + \frac{1}{N} \bigg(\!\!-\frac{\cY\sqrt{\lambda}}{2}+ \ldots \bigg)
        \nonumber\\[2mm]
        & \qquad+ \frac{1}{N^2} \bigg(
        \frac{\cY^2\lambda}{8} + \ldots +\frac{\lambda^{3/2}}{96} + \ldots\bigg) + O(1/N^3)\Bigg]
\end{align}
where for simplicity we only presented the leading contribution arising from the rescaling at each order. Subleading terms can be easily generated. 
Compared with the $\mathcal{N}=4$ result, we observe the appearance of odd powers of $1/N$ and a modification of the $1/N^2$-term. In particular, the new contribution proportional to $\mathcal{Y}^2\lambda$, resulting from the rescaling of the coupling, dominates over the $\lambda^{3/2}$-correction inherited from $\mathcal{N}=4$ SYM. The latter can therefore be neglected at the leading order. We also observe that the $1/N$-terms proportional to $\mathcal{Y}$ generated by the rescaling fully agree with the earlier results of \cite{Beccaria:2021ism} for the conformal \textbf{D}-theory, while they are a new prediction for the non-conformal $\widehat{\mathbf{D}}$-theory.

\paragraph{The function $\widehat{\mathcal{W}}^{(2)}$:} We now consider the contribution $\widehat{\mathcal{W}}^{(2)}$ defined in (\ref{what1}) which starts contributing  at order $1/N^2$.
Indeed, according to (\ref{defRtilde}), we have 
\begin{align}
\label{eq:leaddiag}
        \mathord{
		\begin{tikzpicture}[scale=0.8,baseline=-0.65ex]
			\begin{feynman}
				\vertex (O) at (0,0);
                \vertex (OO) at (1.2,0);
				\vertex (A) at (0,0); 
				\vertex (AA) at (-1,1);
				\vertex (AAA) at (1,1);
				\vertex (B) at (-1.,0) ;
                \vertex (D) at (-0.9,0.3)  {\footnotesize{$i$}};
				\vertex (C) at (1.,0) ;
				\diagram*{
					(O) -- [plain, half left] (OO),
                    (O) -- [plain, half right ] (OO),
						(B) -- [thick,dashed] (O),
				};
			\end{feynman}
			\filldraw[fill=red!20, draw=black] (0,0) circle (0.4cm);
            \begin{feynman}
				\vertex (o) at (0,0) {\scalebox{.65}{$\widehat{R}^{(2)}$}};
			\end{feynman}
	\end{tikzpicture} } &= \frac{1}{2}\,\widehat{R}^{(2)}_{i;j_1 j_2} \big\langle\!\!\big\langle
        K_{j_1} K_{j_{2}}\big\rangle\!\!\big\rangle=\frac{1}{2}\,Q^{(1|2)}_{i;j_1 j_2} \big\langle\!\!\big\langle
        K_{j_1} K_{j_{2}}\big\rangle\!\!\big\rangle+O(1/N^2)\notag
        \\[2mm]
        &=\frac{1}{N}\,\frac{\beta_i \gamma_{j_1} \gamma_{j_2}\, X_{j_1 j_2}}{2} +O(1/N^2)
\end{align} 
where in the final step we used the fact that $Q^{(1|2)}_{i;j_1 j_2}=\frac{\beta_i \gamma_{j_1} \gamma_{j_2}}{N}+O(1/N^3)$ (see Appendix~\ref{sec:GC}) and
replaced the correlator $ \big\langle\!\!\big\langle K_{j_1} K_{j_{2}}\big\rangle\!\!\big\rangle$ with its large-$N$ expansion (\ref{JJ}).
Following the method developed in \cite{Beccaria:2023kbl}, this expression can be further reduced to
\begin{align}
\label{eq:leaddiag1}
        \mathord{
		\begin{tikzpicture}[scale=0.8,baseline=-0.65ex]
			\begin{feynman}
				\vertex (O) at (0,0);
                \vertex (OO) at (1.2,0);
				\vertex (A) at (0,0); 
				\vertex (AA) at (-1,1);
				\vertex (AAA) at (1,1);
				\vertex (B) at (-1.,0) ;
                \vertex (D) at (-0.9,0.3)  {\footnotesize{$i$}};
				\vertex (C) at (1.,0) ;
				\diagram*{
					(O) -- [plain, half left] (OO),
                    (O) -- [plain, half right ] (OO),
						(B) -- [thick,dashed] (O),
				};
			\end{feynman}
			\filldraw[fill=red!20, draw=black] (0,0) circle (0.4cm);
            \begin{feynman}
				\vertex (o) at (0,0) {\scalebox{.65}{$\widehat{R}^{(2)}$}};
			\end{feynman}
	\end{tikzpicture} } &=\frac{1}{N}\,\frac{\beta_i \,w_{0,0}}{8} +O(1/N^2)
\end{align}
where $w_{0,0}$ is a matrix element of the resolvent Bessel operator of the \textbf{E}-theory, already encountered in the analysis of the free energy.

Substituting (\ref{eq:leaddiag1}) into (\ref{what1}), we obtain
\begin{align}
    \widehat{\mathcal{W}}^{(2)}&= \frac{1}{N^2} \frac{w_{0,0}}{8} \sum_{i=1}^\infty \frac{\beta_i}{(2i)!}\Big(\frac{\lambda}{2}\Big)^i + O(1/N^3) \notag\\
    &=\frac{1}{N^2} \frac{w_{0,0}\,\sqrt{\lambda}\,\,I_1\big(\sqrt{\lambda}\big)}{16}+ O(1/N^3)~.
\end{align}
This matches the subplanar term in the expectation value of the Wilson loop of the \textbf{E}-theory computed in \cite{Beccaria:2023kbl}.
Using the asymptotic behavior $w_{0,0} \underset{\lambda \rightarrow \infty}{\sim} -\frac{\sqrt{\lambda}}{2}  + \ldots$ derived in that reference, we finally find  
\begin{align}
    \widehat{\mathcal{W}}^{(2)} \underset{\lambda \rightarrow \infty}{\sim} \frac{1}{N^2}\,\sqrt{\frac{2}{\pi}}\, \frac{\ee^{\sqrt{\lambda}}}{\lambda^{3/4}} \bigg(\!\!-\frac{\lambda^{3/2}}{64}+\ldots\bigg) +O(1/N^3)~.
\end{align}
Adding this contribution to $\widehat{\mathcal{W}}^{(0)}$ modifies the coefficient of the $\lambda^{3/2}$-term in the $\mathcal{N}=4$ SYM expansion (\ref{N4llt}), without affecting the leading term proportional to $\mathcal{Y}^2\lambda$.

This pattern persists also in all other diagrams of (\ref{WtoG1}). Indeed, the dominant strong-coupling contributions of these diagrams are the same as in  the \textbf{E}-theory but they are subleading with respect to those produced by performing the rescaling $\lambda\to\widetilde{\lambda}$ in previous diagrams. Therefore, if we restrict to just the leading strong-coupling contributions at each order in $1/N$, we can neglect $\widehat{\mathcal{W}}^{(2m)}$ with $m\geq1$ and retain only the leading terms from $\widehat{\mathcal{W}}^{(0)}$ given in (\ref{wn4J}). In this way, at leading order the result takes the remarkably simple form
\begin{align}
  \big\langle \cW \big\rangle_\Dhat & ~\underset{\lambda \rightarrow \infty}{\sim}~
        \frac{2\,I_1\big(\!\scaleobj{0.9}{\sqrt{\widetilde\lambda}}\,\big)}{\sqrt{\widetilde\lambda}}+ \ldots = \sqrt{\frac{2}{\pi}}\, \frac{\ee^{\sqrt{\widetilde{\lambda}}}}{\widetilde{\lambda}^{3/4}}\bigg(1 - \frac 38 
        \frac{1}{\sqrt{\widetilde{\lambda}}} + \ldots\bigg) +\ldots
        \label{WLfinal}
\end{align}
with $\widetilde{\lambda}$ given in (\ref{lambdatilde}). For the conformal \textbf{D}-theory this extends the analysis of \cite{Beccaria:2021ism}, while for the $\widehat{\mathbf{D}}$-theory it constitutes a new result, with potential implications for a holographic interpretation.

A few comments are in order. The method we have described allows in principle to compute also the subleading corrections. If $b_0=0$ this can be done systematically along the same lines described in Section~\ref{subsecn:subleading} for the free energy, yielding a generalization of the preliminary results of \cite{Beccaria:2021ism}. However, we have not discussed these calculations since our main focus, and the main novelty of our analysis, is the expectation value of the Wilson loop at large $N$ and large $\lambda$ in the non-conformal case. When $b_0\not=0$ the subleading contributions are more difficult to organize in a compact form and for this reason we only discussed the leading contributions at each order in the $1/N$ expansion. Finally, from a holographic standpoint, the leading behavior $\ee^{\sqrt{\widetilde{\lambda}}}$ in (\ref{WLfinal}) strongly suggests that in the conformal case the dominant string world-sheet contribution should remain the familiar minimal-area one, as in $\mathcal{N}=4$ SYM, but in a set-up where the string tension is related to $\widetilde{\lambda}$ instead of $\lambda$.

\section{Conclusions and outlook}
\label{secn:concl}
In this work we have shown that $\mathcal{N}=2$ SYM theories with two antisymmetric and $N_\fund$ fundamental hypermultiplets can be efficiently studied using supersymmetric localization, even in the presence of a non-vanishing $\beta$-function. The resulting matrix models admit an effective description in terms of gauge-invariant variables whose interactions are controlled by Gaussian multi-trace correlators with a well-defined large-$N$ structure. 

Focusing on the free energy and the $\frac{1}{2}$-BPS Wilson loop, we have shown that the leading strong-coupling effects generated by single-trace interactions can be resummed to all orders into a simple redefinition of the 't Hooft coupling, $\lambda\to\widetilde{\lambda}$.
In the non-conformal $\widehat{\mathbf{D}}$-theory this yields new strong-coupling predictions, while in the conformal \textbf{D}-theory it 
represents a systematic extension of previous results to subleading orders in $1/N$.

Our analysis in the \textbf{D}-theory suggests a refined holographic dictionary in which $\widetilde{\lambda}$ is related to the effective string tension. Using this dictionary, the strong-coupling expansion of the free energy exhibits the characteristic structure expected from a dual string description. 
While the leading and subleading terms admit a natural holographic interpretation, the $g_s^0$-contributions predicted by localization constitute genuinely new data which would be important to confirm with a direct derivation from the gravity side.
Several directions merit further investigation. It would be particularly interesting to extend the strong-coupling expansion to subleading terms in the non-conformal $\widehat{\mathbf{D}}$-theory and clarify the interpretation of the effective coupling $\widetilde{\lambda}$ in that context. 
More generally, our results indicate that localization provides a controlled window into the strong-coupling regime even in non-conformal settings, opening the way to systematic studies of large-$N$ dynamics, resurgent structures, and holographic duals beyond the conformal case. 

Finally, we observe that, in close analogy with the $\widehat{\mathbf{D}}$-theory discussed in this paper, the conformal theory with gauge group Sp($2N$), containing one anti-symmetric and four fundamental hypermultiplets, also admits non conformal extensions in which the number of massless fundamental hypermultiplets is reduced below four. Furthermore, as discussed for example in \cite{Fucito:2023txg,Behan:2024vwg}, starting from the conformal case one can also trigger renormalization group flows ending in strongly-coupled Argyres-Douglas fixed-points. It would be interesting to explore whether a similar pattern arises also for the $\mathbf{D}$-theory with group SU($N$).

We plan to return to some of these issues in future work.

\vskip 1cm
\noindent {\large {\bf Acknowledgments}}
\vskip 0.2cm
We sincerely thank Lorenzo De Lillo, Marialuisa Frau, Francesco Galvagno, Grisha Korchemsky and Paolo Vallarino for many fruitful and enlightening discussions throughout the project. We are also grateful to Arkady Tseytlin and Carlos Barredo Martinez for  valuable comments
and critical remarks which improved the presentation and completeness of the paper. 
This research is partially supported by the INFN project ST\&FI ``String Theory \& Fundamental Interactions''.
The work of A.T. was supported by the French National Agency for Research grant ``Observables'' (ANR-24-CE31-7996).

\vskip 1cm

\appendix 
\section{Gaussian correlators}
\label{sec:GC}
In this section, we summarize a few properties and some new results regarding the connected Gaussian correlators $Q^{(n|2m)}_{i_1,\ldots,i_n;j_1,\ldots,j_{2m}}$ (\ref{defQs}). To begin with, we recall that these correlators   admit the following topological expansion  \cite{Beccaria:2023kbl} in the large-$N$ limit
\begin{align}
Q^{(n|2m)}_{i_1,\ldots,i_n;j_1,\ldots,j_{2m}}&=\frac{\beta_{i_1}\ldots\beta_{i_n}\gamma_{j_1}\ldots\gamma_{j_{2m}}}{N^{n+2m}}\,
\sum_{g=0}^\infty N^{2-2g}\,P^{(n|2m)}_{g}(i_1,\ldots,i_n;j_1,\ldots,j_{2m})\ ,
\label{expansion}
\end{align} 
where the notation was introduced for\,\footnote{Our definitions (\ref{betagamma}) differ by a factor of $\sqrt{2}$ with respect to those in  \,(3.3) of \cite{Beccaria:2023kbl}.}
\begin{align}
\label{betagamma}
\beta_i=\frac{1}{2^{i-1}}\,\frac{\Gamma (2i)}{\Gamma (i)^2}~,\qquad
\gamma_j=\frac{1}{2^{j+\frac{1}{2}}}\,\frac{\Gamma (2j+2)}{\Gamma (j+2)\Gamma (j)}~,
\end{align}
while $P_g^{(n|2m)}$ are symmetric polynomials in $\{i_1,\ldots,i_n\}$ and $\{j_1,\ldots,j_{2m}\}$ of degree
$n+2m+3g-3$.
These polynomials can be conveniently expressed in terms of 
\begin{subequations}
\label{eq:eo}
\begin{align}
e_1&=\sum_{1\leq k\leq n} i_k~,\qquad e_2=\sum_{1\leq k_1<k_2\leq n} i_{k_1}i_{k_2}~,\qquad
\ldots~,\qquad e_n=i_{1}i_{2}\ldots i_n~,\\[1mm]
o_1&=\sum_{1\leq \ell\leq 2m} j_\ell~,\quad~~o_2=\sum_{1\leq \ell_1<\ell_2\leq 2m} j_{\ell_1}j_{\ell_2}~,\quad~
\ldots~,~\quad~\, o_{2m}=j_{1}j_{2}\ldots j_{2m}~.
\end{align}%
\end{subequations}
The set $\{e_k\}$, with $k=1,\ldots,n$, forms a basis for the symmetric polynomials constructed with the $n$ indices of the even operators. Similarly, the set $\{o_\ell\}$, with $\ell=1,\ldots,2m$, is a basis for the symmetric polynomials constructed with the $2m$ indices $j_1,\ldots,j_{2m}$ of the odd  operators. The explicit form of $P^{(n|2m)}_g$ is very involved and not known in general. However, in some cases explicit formulas exist. For example,
one has
\begin{subequations}
    \begin{align}
    P_0^{(n|0)}&=\big(e_1-1\big)_{n-3}~,\label{P0n0}\\[2mm]
    P_0^{(n|2)}&=\big(e_1+o_1)_{n-1}~,\\[1mm]
    P_0^{(n|4)}&=(e_1+o_1)\,\big(e_1+o_1\big)_{n}-\sum_{k=1}^nk!\,e_k(e_1+o_1-k)_{n-k}~,\\
    P_1^{(n|0)}&=\frac{1}{12}\Big[(e_1-7)\,\big(e_1-1)_{n-1}-\sum_{k=2}^n(k-2)!\,e_k\big(e_1-k\big)_{n-k}\,\Big]~,\label{P1n0}
\end{align}
\label{Pgnm}%
\end{subequations}
where the notation $\big(x\big)_k=\frac{\Gamma(x+1)}{\Gamma(x+1-k)}$ stands for the falling factorial. While (\ref{P0n0}) has been known for a long time
\cite{Tutte:1962}, the other genus 0 polynomials $P_0^{(n|2m)}$ have been derived only recently \cite{Bouttier:2024}. To the best of our knowledge, the genus 1 expression (\ref{P1n0}) is instead a new result, which we have been able to infer after computing explicitly $P_1^{(n|0)}$ for many values of $n$ using the
fusion/fission identities of SU($N$). With this method it is not difficult to generate also polynomials with higher values of $g$, $m$ and $n$ case by case, but it is not easy to obtain a simple closed-form expression. Luckily, for the analysis of the first few orders of the large-$N$ expansion only a few of these higher-order polynomials are needed to reconstruct the Gaussian correlators and interaction vertices, and thus it will be enough to have their specific expressions without knowing the general formula.

\paragraph{A useful relation:} 
The operator $\cO_2 = \tr \frac{a^2}{N}$ has a special status since it is proportional to the Gaussian measure. Consequently, the insertion of $\cO_2 $ in a connected correlator can be simplified as follows. Let $\cO_\Delta$ be a product of operators with total dimension $\Delta$. Then
\begin{equation}
    \label{rtecDelta}
        \big\langle\underbrace{\cO_2 \ldots \cO_2} _n \cO_\Delta\big\rangle^{c}_0 = \frac{1}{N^n} \frac{\Gamma(\Delta/2 + n)}{\Gamma(\Delta/2)} 
        \,\big\langle {\cO_\Delta}\big\rangle_0^c~.
\end{equation}
A particular case of this relation, which will be crucial in the following, is
\begin{equation}
    \label{recG}
        Q^{(n+q|2m)}_{1,\ldots,1,i_1,\ldots,i_q;j_1,\ldots j_{2m}} 
        = \frac{1}{N^n} \frac{\Gamma\left(\sum (j_k + \frac 12) + \sum i_l + n\right)}{\Gamma\left(\sum (j_k + \frac 12)+ \sum i_l\right)} Q^{(q|2m)}_{i_1,\ldots, i_q;j_1,\ldots j_{2m}}~. 
\end{equation}
Inserting here the expansion (\ref{expansion}) and taking into account that $\beta_1 = 1$ we deduce that
\begin{align}
    \label{recP}
        P_g^{(n+q|2m)}(1,\ldots,1,i_1,\ldots,i_q;j_1,\ldots, j_{2m}) & = \frac{\Gamma\left(\sum (j_k + \frac 12) + \sum i_l + n\right)}{\Gamma\left(\sum (j_k + \frac 12) + \sum i_l \right)}\nonumber\\
        & \times P_g^{(q|2m)}(i_1,\ldots, i_q;j_1,\ldots, j_{2m})~. 
\end{align}
Thus all even indices equal to 1 can be removed at the price of multiplying by a ratio of $\Gamma$ functions, and this is essential for our manipulations at strong coupling.

\section{Strong-coupling expansions}
\label{App:strong}
In this Appendix, we analyze the strong-coupling behavior of different quantities entering the calculations of the free energy and the  Wilson loop expectation value in the $\widehat{\mathbf{D}}$-theory.

\subsection{Derivation of (\ref{Yastrong})}
\label{App:leading}
In this subsection, we turn our attention to the large-$\lambda$ behavior (\ref{Yastrong}) of the functions $\mathcal{Y}_a $ (\ref{Yell}). For later convenience, we recall that 
   \begin{align}
		\label{eq:Ylapp}
		\mathcal{Y}_{a}
		&=\sum_{m=1}^{\infty}\frac{2(m+1)^{a}}{\Gamma(m+2)^2}\Gamma(2m+2)\zeta({2m+1}) (-1)^m \left(\frac{\lambda}{16\pi^2}\right)^{m+1}\left(4^{m+1}-4+b_0 \right) \ ,
\end{align} where  $b_0$  is given by (\ref{b0Dhat}). To proceed with the calculation, we observe that (\ref{eq:Ylapp}) is given by products of  meromorphic functions that admit well-defined analytic continuations in the complex plane. As a result, we can express $\mathcal{Y}_{a}$ in terms of Mellin-Barnes integrals, which allows us to extract the large-$\lambda$ behavior. To do so, we express  $\mathcal{Y}_{a} $ as 
\begin{equation}
\label{Ysplit}
    \mathcal{Y}_{a} = \mathcal{Y}_{a}^{(1)} +  b_0 \mathcal{Y}_{a}^{(2)} \  .
\end{equation}

In particular, $\mathcal{Y}_{a}^{(1)}$ is independent of the coefficient $b_0$ and takes the following form  \begin{equation}
\label{Y1}
\begin{split}
	\mathcal{Y}_{a}^{(1)}&=2\sum_{m=1}^{\infty}\frac{(m+1)^{a}}{\Gamma(m+2)^2}\Gamma(2m+2)\zeta({2m+1}) (-1)^m \left(\frac{\lambda}{16\pi^2}\right)^{m+1}\left(4^{m+1}-4\right) \\[2mm]
    &=2\int_{\delta-i\infty}^{\delta+i\infty} \dfrac{ds}{2\pi i}\frac{\Gamma(-s)\Gamma(2s+2)(s+1)^{a-1}}{\Gamma(s+2)}\left(\frac{\lambda}{4\pi^2}\right)^{s+1}\eta_{2s+1}\ ,
\end{split}
\end{equation} where the integration contour runs parallel to the imaginary axis with $\delta\in(0,1)$, while  $\eta(s)$ is the Dirichlet $\eta$-function. Closing the contour to the right, we reproduce the perturbative expansion in the first line of (\ref{Y1}). Conversely, closing the contour to the left, we obtain  
\begin{equation}
\label{eq:Y1strong}
	\mathcal{Y}_{a}^{(1)} \underset{\lambda \rightarrow \infty}{\sim} -\frac{\lambda\log 2}{2\pi^2} +\frac{1}{4}\delta_{a,1} \ .
\end{equation} The previous expression holds up to exponentially suppressed terms. 

The second contribution in (\ref{Ysplit}) is proportional to $b_0$ and consequently, it is a specific feature of theories with non-vanishing $\beta$-function. Going through the calculation, we find 
 	\begin{equation}
		\begin{split}
			\mathcal{Y}_{a}^{(2)}&=2\sum_{m=1}^{\infty}(m+1)^{a} \frac{\Gamma(2m+2)}{\Gamma^2(m+2)}\zeta({2m+1})(-1)^m\left(\frac{\lambda}{16\pi^2}\right)^{m+1}\\
			&=2 \int_{\delta-i\infty}^{\delta+i\infty} \dfrac{ds}{2\pi i} \frac{\Gamma(-s)\Gamma(2s+2)(s+1)^{a-2}}{\Gamma(s+1)}\zeta({2s+1})\left(\frac{\lambda}{16\pi^2}\right)^{s+1} \ ,
		\end{split}
	\end{equation}  where the integration contour runs again parallel to imaginary axis with $\delta\in(0,1)$.
    Closing the contour to the left we find, up  to exponentially small corrections, the following result
\begin{equation}
\label{Y2strong}
		\mathcal{Y}_{a}^{(2)}\underset{\lambda \rightarrow \infty}{\sim} -\frac{\lambda\log\lambda}{16\pi^2} -\frac{\lambda}{16\pi^2}\left(2\gamma_E -\log16\pi^2+a\right)-\frac{1}{12}\delta_{a,1}\ .
\end{equation}
Combining together (\ref{eq:Y1strong}) and (\ref{Y2strong}), we find, up to exponentially suppressed corrections, the following result  
\begin{equation}
\label{Yabstrong}
    \begin{split}
        \mathcal{Y}_{a}&\underset{\lambda \rightarrow \infty}{\sim}  -b_0 \frac{\lambda\log\lambda}{16\pi^2}-\frac{\lambda}{16\pi^2}\left(b_0\left(2\gamma_E -\log16\pi^2+a\right)+\log16\right) +\delta_{a,1}\left(\frac{1}{4}-\frac{b_0}{12}\right)\\[2mm]
        &\underset{\lambda \rightarrow \infty}{\sim} -\mathcal{Y}+\ldots  \ ,
    \end{split}
\end{equation} where to obtain the second line we used  the definition of $\mathcal{Y}$  (\ref{Ystrong}).

\subsection{Derivation of (\ref{qellstrong})}
\label{App:subleading}
In this subsection, we use the results obtained in Appendix \ref{App:leading} to derive the large-$\lambda$ behavior (\ref{qellsub}) of the functions $q_\ell$. According to (\ref{eq:qellapp}), we can write
\begin{equation}
    \label{eq:qellapp2}
    \begin{split}
           q_\ell &= -\sum_{g=0}^{\lfloor\frac{\ell-1}{2}\rfloor}\frac{1}{(\ell-2g) !}\prod_{q=1}^{\ell-2g}\left(\beta_{i_q} J_{i_q}\right) P_g^{(\ell-2g|0)}(i_1,\ldots,i_{\ell-2g}) \equiv \sum_{g=0}^{\lfloor\frac{\ell-1}{2}\rfloor}q_{g,\ell} \ ,
    \end{split}
\end{equation} where  implicit summations over the indices $i_1,\ldots$ from $2$ to $\infty$ are understood and we recall that $P^{(\ell-2g|0)}_g$ is a symmetric polynomial of degree $\ell+g-3$ in the variables $i_1,\ldots,i_{\ell-2g}$. The structure of  (\ref{eq:qellapp2}) implies that each term   $q_{g,\ell}$ of the sum in (\ref{eq:qellapp2}) can be expressed as a \emph{homogeneous polynomial} in the variables $\mathcal{Y}_a$ (\ref{eq:Ylapp}) of degree $\ell-2g$. 
For instance, using (\ref{P0n0}) and (\ref{P1n0}), we find the following explicit results 
\begin{subequations}
    \begin{align}
        \label{goeq2}
        q_{0,\ell}
        &= -
        \frac{1}{\ell!} \prod_{q=1}^\ell \big(\beta_{i_q}\,J_{i_q}\big) (e_1 - 1)_{\ell-3} \ ,\\
         \label{g1qell}
        q_{1,\ell} &= -
        \frac{1}{(\ell-2)!} \prod_{q=1}^{\ell-2} \big(\beta_{i_q}\,J_{i_q}\big) \frac{1}{12}\Big[(e_1-7)\,\big(e_1-1)_{\ell-3}-\sum_{k=2}^{\ell-2}(k-2)!\,e_k\big(e_1-k\big)_{\ell-2-k}\,\Big] \ ,
    \end{align}
\end{subequations} where we recall that $e_1$ is defined in (\ref{eq:eo}),  $(x)_k=\frac{\Gamma(x+1)}{\Gamma(x+1-k)}$ denotes the falling factorial. Using these results, it is straightforward to see that both (\ref{goeq2}) and (\ref{g1qell}) can be expressed in terms of the functions $\mathcal{Y}_a(\lambda)$ introduced in (\ref{eq:Ylapp}). Our next task is to study the large-$\lambda$ behavior of these quantities. 

\paragraph{Dominant contributions at large $\lambda$:}
We begin by recalling that the leading contribution to the strong-coupling expansion  (\ref{Yabstrong}) of the functions $\mathcal{Y}_a$ is independent of $a$. As a result, the dominant contribution at large-$\lambda$ to  both (\ref{goeq2}) and (\ref{g1qell}) can be derived by setting  $i_1=\ldots=i_\ell =1$. For instance,  it is straightforward to show that  
\begin{align}
    \label{goeq2dom}
        q_{0,\ell} &\underset{\lambda \rightarrow \infty}{\sim} -
        \frac{1}{\ell!} \prod_{q=1}^\ell \big(\beta_{i_q}\,J_{i_q}\big) (\ell - 1)_{\ell-3}\notag  \\
        &\underset{\lambda \rightarrow \infty}{\sim} -
        \frac{1}{\ell!} \prod_{q=1}^\ell \big(\beta_{i_q}\,J_{i_q}\big) \frac{\Gamma(\ell)}{2} \notag \\
        & \underset{\lambda \rightarrow \infty}{\sim} - \frac{1}{2\ell} \left(-1\right)^\ell \mathcal{Y}^\ell \ ,
\end{align} where to obtain the last line we employed  (\ref{Yabstrong}).  

By similar manipulations,  we also find that 
\begin{align}
    q_{1,\ell} &\underset{\lambda \rightarrow \infty}{\sim}  -
        \frac{1}{(\ell-2)!} \prod_{q=1}^{\ell-2} \frac{\big(\beta_{i_q}\,J_{i_q}\big) }{12}\Big[(\ell-9)\,\big(\ell-3)_{\ell-3}-\sum_{k=2}^{\ell-2}(k-2)! \binom{\ell-2}{k}\big(\ell-2-k\big)_{\ell-2-k}\,\Big]\notag  \\[2mm]
         &\underset{\lambda \rightarrow \infty}{\sim}  -
        \frac{(-1)^\ell\,\mathcal{Y}^{\ell-2}}{12(\ell-2)!}\Big[(\ell-9)\,\Gamma(\ell-2)-\Gamma(\ell-2)(\ell-3)\,\Big]
     =   \frac{(-1)^\ell\,\mathcal{Y}^{\ell-2}}{2(\ell-2)}~,
        \label{eq:q1ell}
\end{align}  
where to obtain the second line we used again (\ref{Yabstrong}). As expected (\ref{eq:q1ell}) scales as $\mathcal{Y}^{\ell-2}$ and consequently, it is subleading compared to (\ref{goeq2dom}). Therefore, we can conclude that  \begin{equation}
\label{eq:qellstronglead}
    q_\ell \underset{\lambda \rightarrow \infty}{\sim} q_{0,\ell} \underset{\lambda \rightarrow \infty}{\sim} - \frac{1}{2\ell} \left(-1\right)^\ell \mathcal{Y}^\ell \ .
\end{equation} The previous expression coincides with (\ref{qellstrong}) and holds in conformal $(b_0=0)$ and non-conformal ($b_0\ne 0$) models. 

\paragraph{Subleading corrections  at large $\lambda$:}
Going beyond the approximation (\ref{eq:qellstronglead}) requires to include subleading corrections to (\ref{goeq2dom}). These  arise from the  corrections of order $\lambda$ in the strong-coupling expansion of the $\mathcal{Y}_a$ functions defined in (\ref{Yabstrong}). When $b_0\ne0$, these terms  depend on $a$ and therefore, the argument we exploited to derive (\ref{goeq2dom}) and (\ref{eq:q1ell}) is invalidated. However, a remarkable simplification occurs when we consider superconformal models. The reason is that  only the function $\mathcal{Y}_1$ develops a subleading $a$-dependent term. For later convenience, we  rewrite  (\ref{eq:Y1strong}) as follows
\begin{equation}
    \label{Yasagain}
        \cY_k \sim -\cY + \nu\, \delta_{k,1}~,
\end{equation}
where $\nu = \frac 14$, while in the conformal case the quantity $\mathcal{Y}$  is explicitly given by $\mathcal{Y}=\frac{\lambda\log 2}{2\pi^2}$.

 Since the subleading contributions in the large-$\lambda$ limit of the functions $q_1$ and $q_2$ were derived in Section \ref{secn:free energy} (see in particular (\ref{q1s}), (\ref{q1strongfull}) and (\ref{q1q2sc})),  we can focus on $q_\ell$ with $\ell\ge 3$. The subleading contributions arising from $q_{1,\ell}$ were derived in (\ref{eq:q1ell}), while those emerging from $q_{0,\ell}$ (\ref{goeq2}) have to be determined. To do so, we begin by observing that (\ref{goeq2}) can be rewritten as follows
\begin{align}
    \label{goeq3}
        q_{0,\ell}
        & = 
        - \frac{1}{\ell!}\prod_{q=1}^\ell \big(\beta_{i_q}\,J_{i_q}\big)
        \sum_{p=1}^{\ell-3} s(\ell-3,p)\, (e_1- 1)^p \nonumber\\
        & = 
        - \frac{1}{\ell!}
        \sum_{p=1}^{\ell-3} s(\ell-3,p) \sum_{r=0}^p (-1)^r \binom{p}{r}\left(i_1+i_2+\ldots+i_\ell\right)^r \prod_{q=1}^\ell \big(\beta_{i_q}\,J_{i_q}\big)~.
\end{align}
In the first line of (\ref{goeq3}) we expressed the falling factorial $(x)_k$ in terms of the Stirling numbers of first kind $s(k,p)$\footnote{The Stirling numbers of the first kind count the number of permutations of $k$ elements which have exactly $p$ cycles.} using 
\begin{equation}
    (x)_k = \sum_{p=1}^k s(k,p) x^p\ .
\end{equation}
To proceed with the calculation, we consider the multinomial expansion
\begin{equation}
    \label{e1rexp}
        e_1^r = (i_1 + \ldots + i_\ell)^r = \sum_{\vec s, |\vec s| = r} \frac{r!}{s_1!\, \ldots s_\ell!}
        i_1^{s_1}\, \ldots i_\ell^{s_\ell}~,
\end{equation}
where $\vec s = (s_1,\ldots s_\ell)$ is a vector of non-negative integers. Requiring that $|\vec s| = \sum_i s_i = r$ 
implies that at most $r$ of its entries are non-zero. 
Since $e_1^r$ is inserted inside a sum over all its indices, we can assume that $\vec s$ takes the specific form $\vec s = (\vec t, \vec 0)$,
where $\vec t =(t_1,\ldots,t_r)$, with $t_1\geq t_2\geq \ldots \geq t_r$,
describes a Young diagram $Y_{\vec t}$ of the permutation group $S_r$.

We must then count the number of vectors $\vec s$ which are equivalent to this choice under permutations of the $\ell$ indices. This number is given by the order $\ell!$ of the permutation group $S_\ell$ divided by the order $C_{\vec t}$ of the stability subgroup of the chosen configuration. The latter can be expressed in terms of the numbers $u_k$ of entries of $\vec t$ which are equal to $k$, with $k=1, \ldots r$.  These can be explicitly determined as
\begin{equation}
    \label{ukis}
        u_k = \sum_{i=1}^r \delta_{t_i,k}~,~~~ k=1,\ldots r~.
\end{equation}
We also introduce 
\begin{equation}
    \label{u0is}
        u_0 = \ell - \sum_{k=1}^r u_k~, 
\end{equation}
which is the number of zero entries in $\vec t$. In terms of these quantities, 
\begin{equation}
    \label{ctis}
        C_{\vec t} = \prod_{k=0}^r (u_k)!~.
\end{equation}Combining everything together, we find that 
\begin{align}
    \label{e1rb}
         \sum_{\{i_1,\ldots,i_\ell \}} \left(i_1+\ldots+i_\ell\right)^r  \prod_{q=1}^\ell \big(\beta_{i_q}\,J_{i_q}\big)
        = \sum_{Y_{\vec t}\in S_r} \frac{\ell!\, \cY_1^{u_0}}{(u_0)!}
        \prod_{k=1}^r \frac{r!\, \cY_{k+1}^{u_k}}{(t_k)!(u_k)!}~,
\end{align}where we recall that $\mathcal{Y}_a$ is defined in (\ref{eq:Ylapp}). The previous expression is valid for arbitrary $\lambda$. This means that it also holds when $\lambda\to\infty$. In this case, as observed before (\ref{goeq2dom}),  
 we can set all the indices $i_1,\ldots$ to $1$. This implies that $e_1 = \ell$ and  (\ref{e1rb}) reduces to
\begin{equation}
    \label{e1yall}
        \ell^r (-\cY)^\ell = \sum_{Y_{\vec t}\in S_r}\frac{\ell!\, r!}{(u_0)!\prod_{k=1}^r (t_k)! (u_k)!} (-\cY)^{u_0 + \sum_{k=1}^r u_k}
        = \sum_{Y_{\vec t}\in S_r}\frac{\ell!\, r!}{(u_0)!\prod_{k=1}^r (t_k)! (u_k)!} (-\cY)^\ell~,
\end{equation}
where in the last step we used  (\ref{u0is}). This gives us the identity 
\begin{equation}
    \label{sumYis}
        \sum_{Y_{\vec t}\in S_r}\frac{\ell!\, r!}{(u_0)!\prod_{k=1}^r (t_k)! (u_k)!} = \ell^r~.
\end{equation}

Substituting  (\ref{Yasagain}) into (\ref{e1rb}), it is straightforward to show that 
\begin{align}
    \label{e1rb2_0}
        e_1^r  \prod_{q=1}^\ell \big(\beta_{i_q}\,J_{i_q}\big) 
        & = \sum_{Y_{\vec t}\in S_r}\frac{\ell!\, r!}{(u_0)!}
        (-\cY+\nu)^{u_0} \prod_{k=1}^r \frac{(-\cY)^{u_k}}{(t_k)! (u_k)!}
        \nonumber\\
        & = \sum_{Y_{\vec t}\in S_r}\frac{\ell!\, r!}{(u_0)!} 
        \sum_{j=0}^{\ell} \binom{u_0}{j}\nu^j (-\cY)^{u_0 + \sum_k u_k - j}
        \prod_{k=1}^r \frac{1}{(t_k)! (u_k)!}
        \nonumber\\
        & = \sum_{j=0}^{\ell} \frac{\nu^j}{j!} (-\cY)^{\ell - j}  \frac{\ell!}{(\ell-j)!}
        \sum_{Y_{\vec t}\in S_r}\frac{(\ell-j)!\, r!}{(u_0-j)!} \prod_{k=1}^r \frac{1}{(t_k)! (u_k)!}~, 
\end{align}
where we took into account (\ref{u0is}). The sum over the tableaux can be done using the identity (\ref{sumYis}) with $\ell$ replaced by $\ell-j$. This leads to
\begin{align}
    \label{e1rb2}
        e_1^r  \prod_{q=1}^\ell \big(\beta_{i_q}\,J_{i_q}\big) 
        & = \sum_{j=0}^{u_0} \frac{\ell^j}{j!} (-\cY)^{\ell - j}  \frac{\ell!}{(\ell-j)!} (\ell-j)^r~.
\end{align}
Inserting the previous expression  into  (\ref{goeq3}), we obtain
\begin{align}
    \label{q0nis2}
        q_{0,\ell} & = - \sum_{j=0}^{\ell} \frac{\nu^j}{j!} (-\cY)^{\ell- j}  
        \sum_{p=1}^{\ell-3} s(\ell-3,p) \sum_{r=0}^p (-1)^r \binom{p}{r} (\ell-j)^r
        \nonumber\\
        & = - \sum_{j=0}^{\ell} \frac{\nu^j}{j!} \frac{(-\cY)^{\ell - j}}{\ell-j} \frac{1}{\Gamma(3-j)}.
\end{align}
Note that on the right-hand side only the terms $j=0,1,2$ contribute.  Altogether in the large-$\lambda$ regime, up to constants and exponentially suppressed terms, we find that 
\begin{equation}
    \label{q0nfin}
        q_{0,\ell} \underset{\lambda \rightarrow \infty}{\sim} -(-1)^\ell \frac{\cY^\ell}{2\ell} + (-1)^\ell \frac{\cY^{\ell-1}}{4(\ell-1)} - (-1)^\ell \frac{\cY^{\ell-2}}{32(\ell-2)}~,
\end{equation} where we used the fact that $\nu = 1/4$. The previous expression holds for $\ell\ge3$. Finally, we combine the previous expression with (\ref{eq:q1ell}) and we obtain that for $\ell\ge 3$
\begin{align}
    q_{\ell}&\underset{\lambda \rightarrow \infty}{\sim} q_{0,\ell}+q_{1,\ell}+\ldots \notag \\
    &\underset{\lambda \rightarrow \infty}{\sim} -(-1)^\ell \frac{\cY^\ell}{2\ell} + (-1)^\ell \frac{\cY^{\ell-1}}{4(\ell-1)} - (-1)^\ell \frac{\cY^{\ell-2}}{32(\ell-2)}+  
        \frac{(-1)^\ell}{(\ell-2)}\mathcal{Y}^{\ell-2}\frac{1}{2}\notag \\
        &\underset{\lambda \rightarrow \infty}{\sim}~-(-1)^{\ell}\,\frac{\mathcal{Y}^\ell}{2\ell}+(-1)^{\ell}\frac{\mathcal{Y}^{\ell-1}}{4(\ell-1)}
        +(-1)^{\ell}\frac{15\,\mathcal{Y}^{\ell-2}}{32(\ell-2)}~.
        \label{q0nfinal}
\end{align}
The previous expression coincides with (\ref{eq:qellsub}). There could be further terms of order $\cY^{\ell-2 g}$ with $g\geq 2$ which we are not able to discuss in a closed form because we lack the explicit expression of the polynomials $P_g^{(\ell-2g|0)}$ of higher genera. We can, however, easily compute the explicit expressions of specific connected correlators and we checked in many examples that such potential corrections to $q_\ell$ always vanish.

\subsection{Derivation of (\ref{hatQ2m})}
\label{sec:properties}
The argument we exploited to derive (\ref{goeq2dom}) applies also to $\widehat Q^{(2m)}$ and $\widehat R^{(2m)}$.

Let us begin by considering $\widehat Q^{(2m)}$. From the definition (\ref{Qhat}), we have 
\begin{align}
    \widehat{Q}^{(2m)}_{j_1,\ldots,j_{2m}}&=\sum_{n=0}^\infty \frac{1}{n!}\,J_{i_1}\ldots J_{i_n}\,Q^{(n|2m)}_{i_1\ldots i_n;j_1\ldots j_{2m}} \notag \\
    &= \sum_{n=0}^\infty \frac{1}{n!}\,J_{i_1}\ldots J_{i_n}\frac{\beta_{i_1}\ldots\beta_{i_n}\gamma_{j_1}\ldots\gamma_{j_{2m}}}{N^{n+2m}}
\sum_{g=0}^\infty N^{2-2g}\,P^{(n|2m)}_{g}(i_1,\ldots,i_n;j_1,\ldots,j_{2m})
\label{eq:Qhat2mApp}
\end{align} 
where to obtain the second line we replaced the connected Gaussian correlators $Q^{(n|2m)}$ with their topological expansion (\ref{expansion}) in the large-$N$ limit. As a result, $\widehat{Q}^{(2m)}$ can be expressed in terms of the function $\mathcal{Y}_a$ (\ref{eq:Ylapp}) whose large-$\lambda$ limit is independent of $a$. Therefore, we can follow the argument we employed to derive (\ref{goeq2dom}) and set the indices $i_1,\ldots,i_n$ to 1 in the argument of the symmetric polynomials $P^{(n|2m)}_{g}(i_1,\ldots,i_n;j_1,\ldots,j_{2m})$. In this way we remain with 
\begin{align}
        \widehat Q^{(2m)}_{j_1,\ldots, j_{2m}} &\underset{\lambda \rightarrow \infty}{\sim}~  \sum_{n=0}^\infty \frac{1}{n!} \prod_{q=1}^n \beta_{i_q} J_{i_q} \frac{1}{N^{2m}}\notag
        \sum_{g=0}^\infty N^{2-2g} P_g^{(n|2m)}(1,\ldots,1;j_1,\ldots,j_{2m}) \gamma_{j_1} \ldots \gamma_{j_{2m}}\\
        &\underset{\lambda \rightarrow \infty}{\sim}~  \sum_{n=0}^\infty \frac{1}{n!} \frac{(-\cY)^n}{N^{n+2m}}
        \sum_{g=0}^\infty N^{2-2g} P_g^{(n|2m)}(1,\ldots,1;j_1,\ldots,j_{2m}) \gamma_{j_1} \ldots \gamma_{j_{2m}} \label{Qt2Pll}
\end{align} 
where to obtain the second line we took into account the large-$\lambda$ behavior (\ref{Yabstrong}). Finally, we apply  (\ref{recP}) with $q=0$  and we find that~\footnote{To perform the sum over $n$ in (\ref{Qt2Pllbis}), we exploited 
\begin{equation*}
        \sum_{n=0}^\infty \frac{1}{n!} \frac{\Gamma(M+n)}{\Gamma(M)} x^n = \frac{1}{(1-x)^M} ~.
\end{equation*}
}
\begin{align}
        \widehat Q^{(2m)}_{j_1,\ldots, j_{2m}}&\underset{\lambda \rightarrow \infty}{\sim}~  \sum_{n=0}^\infty \frac{1}{n!} 
        \frac{\Gamma(\sum (j_k +\frac 12)+ n)}{\Gamma(\sum (j_k + \frac 12))}\frac{(-\cY)^n}{N^{n+2m}}
        \sum_{g=0}^\infty N^{2-2g} P_g^{(0|2m)}(j_1,\ldots,j_{2m}) 
        \gamma_{j_1} \ldots \gamma_{j_{2m}}\notag\\
        &\underset{\lambda \rightarrow \infty}{\sim}   \frac{1}{\left(1 + \frac{\cY}{N}\right)^{\sum (j_k + \frac 12)}}\, Q^{(0|2m)}_{j_1,\ldots, j_{2m}}~,\label{Qt2Pllbis}
\end{align} 
where to obtain the final equality we recognized the topological expansion of the Gaussian correlator $Q^{(0|2m)}$ by applying (\ref{expansion}) with $n=0$. Note that (\ref{Qt2Pllbis}) coincides with (\ref{hatQ2m}).

The analysis of the quantities $\widehat R^{(2m)}$ goes along the same lines as their structure closely resembles that of (\ref{eq:Qhat2mApp}). Indeed, starting from their definition (\ref{defRtilde}), we find
\begin{align}
 \widehat R^{(2m)}_{i;j_1\ldots j_{2m}}   & =\sum_{p=0}^\infty \frac{1}{p!} \,J_{k_1}\ldots J_{k_p}\,Q^{(p+1|2m)}_{i,k_1,\ldots, k_p;j_1,\ldots,j_{2m}}\\
 &=\beta_i  \sum_{g,p=0}^\infty \frac{1}{p!}\,J_{k_1}\ldots J_{k_p}\frac{\beta_{k_1}\ldots\beta_{k_p}\gamma_{j_1}\ldots\gamma_{j_{2m}}}{N^{p+1+2m+2g-2}}\,
P^{(p+1|2m)}_{g}(i,k_1,\ldots,k_p;j_1,\ldots,j_{2m})\notag
\end{align}
where to obtain the second line we replaced the connected Gaussian correlators $Q^{(p+1|2m)}$ with their topological expansion (\ref{expansion}).  Following the same approach that was used to derive (\ref{Qt2Pllbis}), we finally obtain
\begin{equation}
    \label{RtoQ12m}
         \widehat{R}^{(2m)}_{i;j_1,\ldots,j_{2m}} ~\underset{\lambda \rightarrow \infty}{\sim}~
         \frac{1}{\left(1 + \frac{\cY}{N}\right)^{\sum (j_k + \frac 12)+ i}}\, Q^{(1|2m)}_{i;j_1,\ldots, j_{2m}}~.
\end{equation}

\begin{comment}

\end{comment}

\bibliographystyle{JHEP}
\bibliography{main}
\end{document}